\documentclass{amsproc}
\pdfoutput=1 
\usepackage{amssymb,amsmath,amscd,bbm,tikz,graphicx}
\input epsf.sty
\usepackage{epic,eepic,epsfig}
\usepackage[all]{xy}
\usepackage{url}
\usepackage[bookmarks=true,%
    colorlinks=true,%
    linkcolor=blue,%
    citecolor=blue,%
    filecolor=blue,%
    menucolor=blue,%
    urlcolor=blue,%
    breaklinks=true]{hyperref}
\usepackage{slashed}    

\newtheorem{theorem}{Theorem}[section]
\theoremstyle{definition}
\newtheorem{proposition}[theorem]{Proposition}

\newtheorem{conjecture}[theorem]{Conjecture}

\newtheorem{example}[theorem]{Example}

\def\tr{{\rm Tr}}

\def\BZ{\mathbbm Z}
\def\IP{\mathbbm P}
\def\IF{\mathbbm F}
\def\IC{\mathbbm C}
\def\IR{\mathbbm R}
\def\IZ{\mathbbm Z}

\def\CO{\mathcal{O}}
\def\CJ{\mathcal{J}}
\def\CF{\mathcal{F}}
\def\CR{\mathcal{R}}
\def\CB{\mathcal{B}}

\newcommand{\fad}{\operatorname{\Phi}_{\mathsf{b}}}

\newcommand{\mypsi}[2]{\operatorname{\Psi}_{#1,#2}}

\newcommand{\be}{\begin{equation}}
\newcommand{\ee}{\end{equation}}
\newcommand{\ba}{\begin{aligned}}
\newcommand{\ea}{\end{aligned}}
\newcommand{\mJ}{{\mathsf{J}}}
\newcommand{\mx}{{\mathsf{x}}}
\newcommand{\my}{{\mathsf{y}}}
\newcommand{\mX}{{\mathsf{X}}}
\newcommand{\mY}{{\mathsf{Y}}}
\newcommand{\mH}{{\mathsf{H}}}
\newcommand{\mb}{{\mathsf{b}}}
\newcommand{\map}{{\mathsf{p}}}
\newcommand{\mq}{{\mathsf{q}}}

\newcommand{\mO}{{\mathsf{O}}}
\newcommand{\re}{{\rm e}}
\newcommand{\ri}{{\rm i}}
\newcommand{\rd}{{\rm d}}
\newcommand{\im}{\mathsf{i}}
\newcommand{\figref}[1]{Fig.~\protect\ref{#1}}

\begin{document}
\title{Spectral theory and mirror symmetry}
\author{Marcos Mari\~no}
\address{Section de Math\'ematiques et D\'epartement de Physique Th\'eorique\\
Universit\'e de Gen\`eve, 1211 Gen\`eve 4, Switzerland }
\email{marcos.marino@unige.ch}
\thanks{
M.M. is supported in part by the Swiss National Science Foundation, subsidies 200020-149226, 200021-156995, 
and by the NCCR 51NF40-141869 ``The Mathematics of Physics" (SwissMAP)}

\begin{abstract}
Recent developments in string theory have revealed a surprising connection between spectral theory and local mirror symmetry: it has been found that 
the quantization of mirror curves to toric Calabi--Yau threefolds leads to trace class operators, whose spectral properties are conjecturally encoded in the enumerative 
geometry of the Calabi--Yau. This leads to a new, infinite family of solvable spectral problems: the Fredholm determinants of these 
operators can be found explicitly in terms of Gromov--Witten invariants and their refinements; their spectrum is encoded in exact quantization conditions, and turns out to be determined 
by the vanishing of a quantum theta function. Conversely, the spectral theory of these operators provides a non-perturbative definition 
of topological string theory on toric Calabi--Yau threefolds. In particular, their integral kernels lead to matrix integral representations of the topological string partition function, which 
explain some number-theoretic properties of the periods. In this paper we give 
a pedagogical overview of these developments with a focus on their mathematical implications. \end{abstract}

\maketitle



\section{Introduction}
\label{sec.intro}


Mirror symmetry has played a fundamental r\^ole in the interface of theoretical physics and mathematics, and has led to beautiful developments in string theory and 
enumerative geometry. Although mirror symmetry was originally formulated for compact Calabi--Yau (CY) 
manifolds, it can be extended to the so-called local case \cite{KKV,CKYZ}, which involves toric, hence non-compact CY manifolds. Local 
mirror symmetry is in a sense even richer than its compact counterpart, since it is related to many other fields of mathematical physics, like Chern--Simons theory, 
integrable systems, supersymmetric gauge theory, and random matrix theory. Very often, these connections have been obtained by studying topological string theory, the physical theory underlying mirror symmetry. 
In some circumstances, this theory can be described by other, ``dual" theories, which look in principle very different, and this physical equivalence leads to a non-trivial 
mathematical equivalence. For example, the large $N$ duality obtained by Gopakumar and Vafa in \cite{gv} relates topological string theory on the resolved conifold 
to Chern--Simons theory on the three-sphere, and this equivalence eventually led to the theory of the topological vertex \cite{AKMV}, which gives a complete algorithm 
to obtain the Gromov--Witten invariants of toric CY manifolds in terms of Chern--Simons knot invariants. 

The spectral theory of self-adjoint operators plays a more established r\^ole in physics, due to its connection to Quantum Mechanics. In recent years, there have been many interesting 
developments in the arena of exactly solvable spectral problems. For example, in some one-dimensional models, exact quantization conditions incorporating non-perturbative 
effects at all orders have been found (see for example \cite{ZJJ,voros2,alvarez}). Some of these solvable models are closely related to integrable systems \cite{DT} and two-dimensional CFT \cite{blz}. 

The purpose of this review paper is to give a summary of some recent results in string theory which suggest a deep connection between these two separate subjects. There is now 
growing evidence that, given a toric CY manifold, one can associate to it trace class operators whose spectral properties are encoded in the enumerative invariants  
of the underlying manifold. This correspondence is quite precise and leads to conjectural, explicit formulae for the Fredholm determinants of these operators, from which one can 
derive exact quantization conditions for their spectrum. From the point of view of spectral theory, this connection provides a new family of solvable operators with beautiful properties. From the point of view 
of enumerative geometry, it gives a new meaning to the generating functionals encoding the enumerative invariants of local CY geometries: they appear as asymptotic expansions of 
quantities defined by the spectral theory of the corresponding quantum operators. For example, one can obtain the genus expansion of the topological string free energy as an asymptotic expansion of the ``fermionic" spectral 
traces of these operators. 

The correspondence between spectral theory and topological strings which we will describe can be regarded in many ways as a large $N$ string/gauge duality, like those appearing in the 
AdS/CFT correspondence. It relates a simple quantum-mechanical problem on the real line to a string propagating on a toric CY manifold. The weak coupling regime of the quantum 
problem, when the Planck constant is small, corresponds to the regime in which the string coupling constant is strong. The genus expansion of the topological string 
emerges in a 't Hooft-like limit of the quantum mechanical problem, which can be regarded as a rigorous non-perturbative definition of topological string theory on 
these backgrounds. 

This review paper is organized as follows. In section \ref{problem}, we present a simple example of the general type of spectral problems 
we will deal with, and we present both the approximate solution obtained with elementary methods and the exact solution conjectured 
in the recent literature. In section \ref{st-ts-sec} we present the general conjecture formulated in \cite{GHM} connecting spectral theory and topological 
string theory on toric CY manifolds. The emphasis is on how string theory solves the spectral problem. In section \ref{ts-st} we explain how, reciprocally, 
the spectral problem leads to new insights on topological string theory, providing in particular a non-perturbative definition of the genus expansion. 
Finally, in section \ref{out}, we conclude with some interesting problems for the future. 

\subsection*{Acknowledgments}
 I would like to thank, first of all, my co-workers on this subject, for the enjoyable collaboration: 
 Santiago Codesido, Ricardo Couso, 
 Alba Grassi, Jie Gu, Yasuyuki Hatsuda, Johan Kallen, Rinat Kashaev, Albrecht Klemm, Sanefumi Moriyama, Kazumi Okuyama, Jonas Reuter, Ricardo Schiappa, and 
 Szabolcs Zakany. In particular, Santiago Codesido, Ricardo Couso, Alba Grassi, Yasuyuki Hatsuda, Rinat Kashaev, Ricardo Schiappa and 
 Szabolcs Zakany read a preliminary version of this paper and made many 
 useful comments. Finally, I would like to thank Boris Pioline and the rest of the organizers of {\it String-Math 2016} for the invitation to 
 speak and for the opportunity to contribute to the proceedings. 

\section{A problem in spectral theory}
\label{problem}

\subsection{A quantum curve}

Let us start by formulating a sharp question concerning the spectrum of an interesting operator on the real line. 
This will hopefully provide a motivation for the type of problems that we would like to address in this overview paper. 

Let $\mx$, $\my$ be Heisenberg operators on the real line, satisfying the commutation relation 
\be
\label{hcr}
[\mx, \my] =\im \hbar. 
\ee
Let us consider the following operator, 
\be
\label{op-p2}
\mO_{\IP^2}= \re^{\mx} + \re^{\my}+ \re^{-\mx -\my}. 
\ee
Why look at this beast? The reason is that it can be obtained by ``quantizing" the mirror curve to the CY threefold known as ``local $\IP^2$", which is the total space of the canonical 
bundle of $\IP^2$:
\be
\label{Xlocalp2}
X=\CO(-3) \rightarrow \IP^2. 
\ee
This is probably the most studied example among all non-compact CY manifolds. As we will briefly review later on, 
the mirror of this manifold is encoded in an elliptic curve in $\IC^\star \times \IC^\star$, which can be written in the form 
\be
\label{mp2}
\re^x + \re^{y} + \re^{-x-y}+\kappa=0. 
\ee
Here, $\kappa$ parametrizes the moduli space of complex structures of the curve. We can now perform the standard Weyl quantization of the first three terms 
of the curve (\ref{mp2}), i.e. we promote $x$ and $y$ to Heisenberg operators and we use the Weyl ordering prescription. This gives the operator (\ref{op-p2}). 
It involves the exponentiated ``coordinate" and ``momentum" operators
\be
\mX= \re^{\mx} , \qquad \mY=\re^{\my},  
\ee
which are self-adjoint and they satisfy the Weyl algebra
\be
\mX \mY= q  \mY \mX, 
\ee
where
\be
\label{q-def} q=\re^{\ri \hbar}. 
\ee
The domain of the operator $\mX$, $D(\mX)$, consists of functions $\psi(x) \in L^2(\IR)$ such that
\be
\re^x \psi(x) \in L^2(\IR).  
\ee
Similarly, the domain of $\mY$, $D(\mY)$, consists of functions $\psi(x) \in L^2(\IR)$ such that 
\be
\label{p-cond}
\re^y \widehat \psi(y) \in L^2(\IR), 
\ee
where 
\be
\widehat \psi(y)=\int {\rd x \over {\sqrt{2 \pi \hbar}}} \re^{-\im xy/\hbar} \psi(x)
\ee
is the wavefunction in the $y$ representation, which is essentially given by a Fourier transform. The condition (\ref{p-cond}) can be translated 
into a condition on $\psi(x)$, requiring analyticity on strips. The domain of the operator $\mO_{\IP^2}$ should be chosen appropriately, by considering functions 
such that $\mO_{\IP^2} \psi \in L^2 (\IR)$. 

As we will review in the next section, it was shown in \cite{KM} that the operator $\mO_{\IP^2}$ has 
an inverse which is positive-definite and of trace class in $L^2(\IR)$, so its spectrum (which depends on $\hbar$) consists of an infinite set of positive eigenvalues 
$\re^{E_n}$, $n=0, 1, \cdots$. This spectrum can be determined numerically with standard methods, as first pointed out in \cite{HW}. First, one 
should choose an appropriate orthonormal basis of functions $\{ \phi_i\}_{i=0,1, \cdots}$ of $L^2(\IR)$, in the domain of $\mO_{\IP^2}$ (for example, the eigenfunctions of  the harmonic 
oscillator will do). Then, the $\re^{E_n}$ are the eigenvalues of the infinite-dimensional matrix 
\be
M_{ij}= (\phi_i, \mO_{\IP^2} \phi_j), \qquad i,j=0,1, \cdots. 
\ee
They can be obtained numerically by truncating the matrix to very large sizes. Boundary effects can be partially eliminated with standard extrapolation methods, 
and one can find very accurate values for the $E_n$. For $\hbar=2 \pi$, 
the results for the very first eigenvalues are listed in Table \ref{table-p2}. We can now ask the following question: can these eigenvalues be determined {\it analytically}? 
Usually, this is hopeless, since there are very few operators whose spectral properties can be determined exactly. However, a conjecture put forward in \cite{GHM} affirms 
that in this case it can be done: there is an exact, conjectural expression for the Fredholm determinant of the inverse operator 
\be
\label{rho-simple}
\rho_{\IP_2}= \mO_{\IP^2}^{-1},
\ee
from which one can deduce an {\it exact quantization condition} for the spectrum\footnote{The exact quantization condition for this particular operator 
and $\hbar=2 \pi$ was first written down in \cite{HW} by 
generalizing a result of \cite{KaMa} for a different CY geometry.}. Before presenting this conjecture, let us see what can be said about the spectrum of the above operator with elementary methods.   
\begin{table}[t] 
\centering
\begin{tabular}{l  l}
\hline
$n$ &	$E_n$  \\
\hline
   0&  2.56264206862381937\\
 1 &   3.91821318829983977\\
   2&  4.91178982376733606\\
    3 & 5.73573703542155946\\
   4&  6.45535922844299896\\    
\hline
\end{tabular}
\vskip .25cm
\caption{Numerical spectrum of the operator (\ref{op-p2}) for $n=0, 1, \cdots, 4$, and $\hbar=2 \pi$.}
\label{table-p2}
\end{table}

\subsection{A first approach to the spectrum}

Although the above spectral problem is not of the standard form found in Quantum Mechanics, it is possible to use conventional 
approximation techniques to gain some insight. In particular, 
one can try to use the WKB approximation. Let us briefly review this method. 
Let us assume that we have a classical Hamiltonian of the form, 
\be
\label{HCM}
H(x,y)= {y^2 \over 2} + V(x), 
\ee
where $y$ is interpreted as the momentum, and $V(x)$ is a potential supporting bound states. 
We will assume for simplicity that $V(x)$ is a confining potential, i.e. that $V(x) \rightarrow +\infty$ for $|x| \rightarrow \infty$. 
After quantization, one finds a quantum Hamiltonian $\mH$ with a discrete spectrum of energies $E_n$, $n=0, 1, \cdots$. The Bohr--Sommerfeld quantization 
condition gives an {\it approximate} quantization condition which determines the spectrum in the limit of small $\hbar$ or large quantum numbers $n \gg1$. 
It can be formulated geometrically as follows. One considers the plane curve 
\be
\label{HE}
H(x,y)=E,
\ee
where $E$ is interpreted as the energy of the system, and a differential on this curve given by 
\be
\label{diff}
\lambda= y(x) \rd x. 
\ee
In simple situations, there is a $B$-cycle in the curve (\ref{HE}) which corresponds to periodic motion in the potential, between two turning points determined by the 
equation of the curve. The integral of $\lambda$ over this cycle gives the volume of the region $\CR(E)$ in phase space with energy less or equal than $E$, 
\be
\label{cr}
\CR(E)=\{ (x,y) \in \IR^2: H(x,y) \le E \}.
\ee
We will denote this volume by ${\rm vol}_0(E)$. The Bohr--Sommerfeld quantization condition reads
\be
\label{BS}
{\rm vol}_0(E) = \oint_B \lambda = 2\pi \hbar \left( n+{1\over 2} \right), \qquad n=0, 1, \cdots. 
\ee
It can be also interpreted as saying that each cell of volume $2 \pi \hbar$ in $\CR(E)$ corresponds roughly to a quantum state. 

\begin{figure}[h]
\center
\includegraphics[scale=0.35]{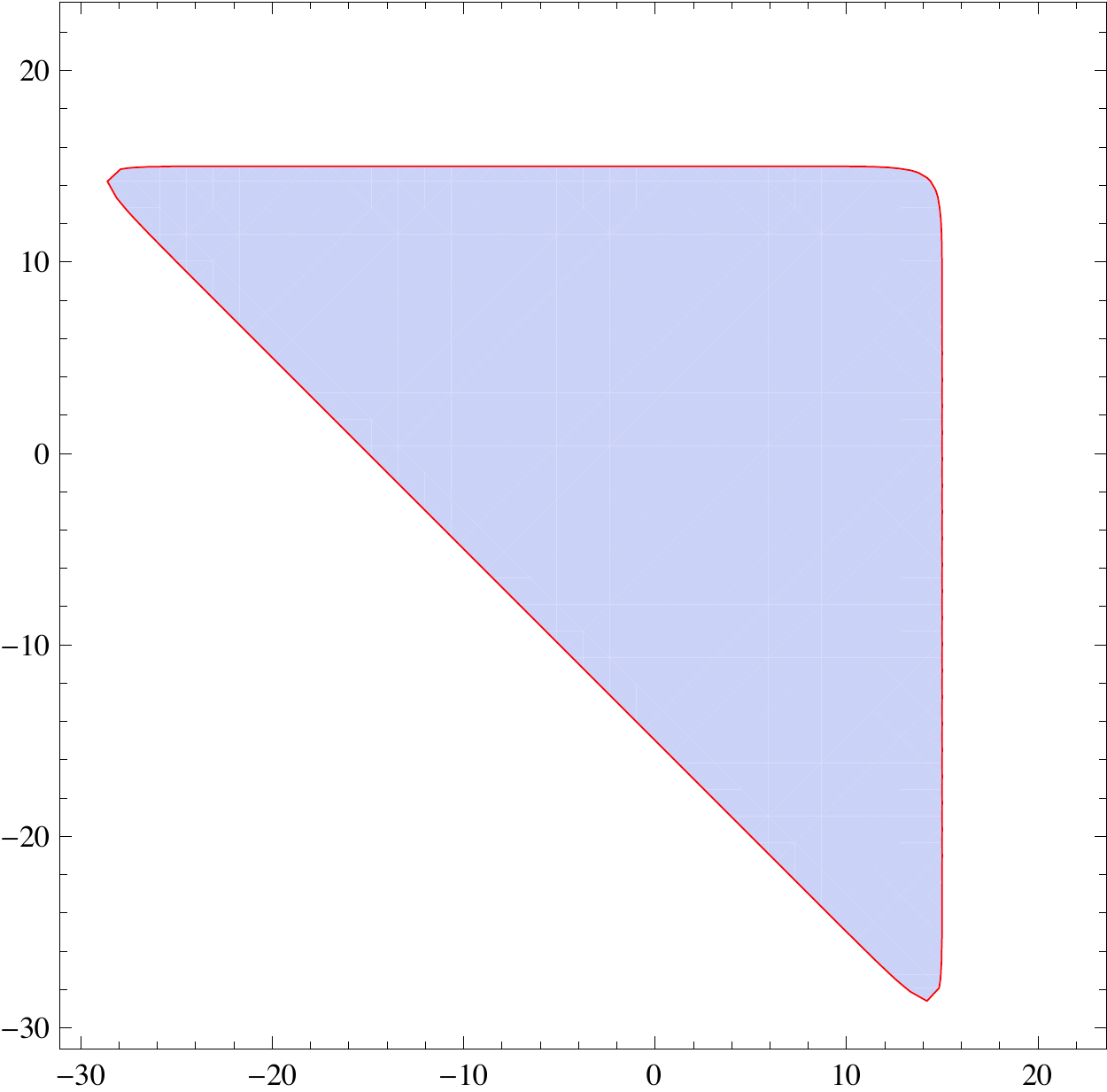}
\caption{The region $\CR(E)$ for $E=15$.}
\label{p2-reg}
\end{figure}

It turns out that, although our quantum operator (\ref{op-p2}) is not of the standard form (\ref{HCM}), we can still use the Bohr--Sommerfeld condition. 
The classical counterpart of the operator $\mO_{\IP^2}$ 
is the function on the phase space $\IR^2$ given by the first three terms in (\ref{mp2}), with the 
symplectic form $\omega = \rd x \wedge \rd y$, 
\be
\label{op2-f}
\CO_{\IP^2}(x,y)= \re^x + \re^y + \re^{-x-y}. 
\ee
The analogue of the hyperelliptic curve (\ref{HE}) is now, 
\be
\CO_{\IP^2}(x, y) =\re^E, 
\ee
and the function (\ref{op2-f}) defines a compact region in phase space 
\be
\label{re}
\CR(E)=\{ (x,y)\in \IR^2: \CO_{\IP^2}(x, y) \le \re^E\}, 
\ee
which is the analogue of (\ref{cr}). Note that we are implicitly interpreting $\CO_{\IP^2}$ as the exponential of a classical Hamiltonian. 
The Bohr--Sommerfeld quantization condition for this problem is again given by (\ref{BS}). The period of $\lambda$ can be computed 
explicitly, but since the resulting quantization condition is only valid at large energies, it is worth to simplify it, as follows. 
At large $E$, the region $\CR(E)$ has a natural mathematical interpretation: it is the region enclosed by the 
tropical limit \cite{grisha} of the curve 
\be
\re^x + \re^y + \re^{-x-y}-\re^E=0, 
\ee
and limited by the lines 
\be
x=E, \qquad y=E, \qquad x+y+E=0, 
\ee
see \figref{p2-reg} for an illustration when $E=15$. In this polygonal limit, we find 
\be
{\rm vol}_0(E) \approx { 9 E^2 \over 2}, 
\ee
and we obtain from (\ref{BS}) the following approximate behavior for the eigenvalues, at large quantum numbers, 
\be
\label{bs}
E_n \approx {2\over 3} {\sqrt{\pi \hbar}}\,  n^{1/2}, \qquad  n\gg 1. 
\ee
It can be seen, by direct examination of the numerical spectrum, that this rough WKB estimate gives a good approximation to the eigenvalues of the operator (\ref{op-p2}) when 
$n$ is large. The estimate (\ref{bs}) has been rigorously proved in \cite{LST}. 

\subsection{An exact quantization condition}

An important question is whether the Bohr--Sommerfeld quantization condition obtained above can be improved. One possible strategy is to consider 
the WKB quantization scheme to {\it all orders} in $\hbar$. The all-orders WKB quantization condition was derived for standard Schr\"odinger operators 
by Dunham in \cite{D}. It can be adapted to operators of the form (\ref{rho-simple}) by using a general formalism due to Voros \cite{voros1}, or by 
using the extension of the WKB method to difference equations in \cite{dingle}. 
For the operator (\ref{op-p2}), this all-orders WKB method was studied in \cite{ACDKV}. As we will explain in detail in the next section, 
the main (conjectural) result of \cite{ACDKV} is that the resulting quantization condition can be expressed in terms of enumerative invariants of the CY 
(the so-called NS limit of the refined topological string.) However, as pointed out in \cite{KaMa}, this 
can not be the full answer to the spectral problem, since it leads to unacceptable 
poles for values of $\hbar$ of the form $\hbar=q \pi$, where $q\in \mathbb{Q}$. Physically, one has to 
take into account the contribution from quantum-mechanical instantons, which are known 
to correct the all-orders WKB quantization condition \cite{bpv}. 

\begin{figure}[h]
\center
\includegraphics[scale=0.6]{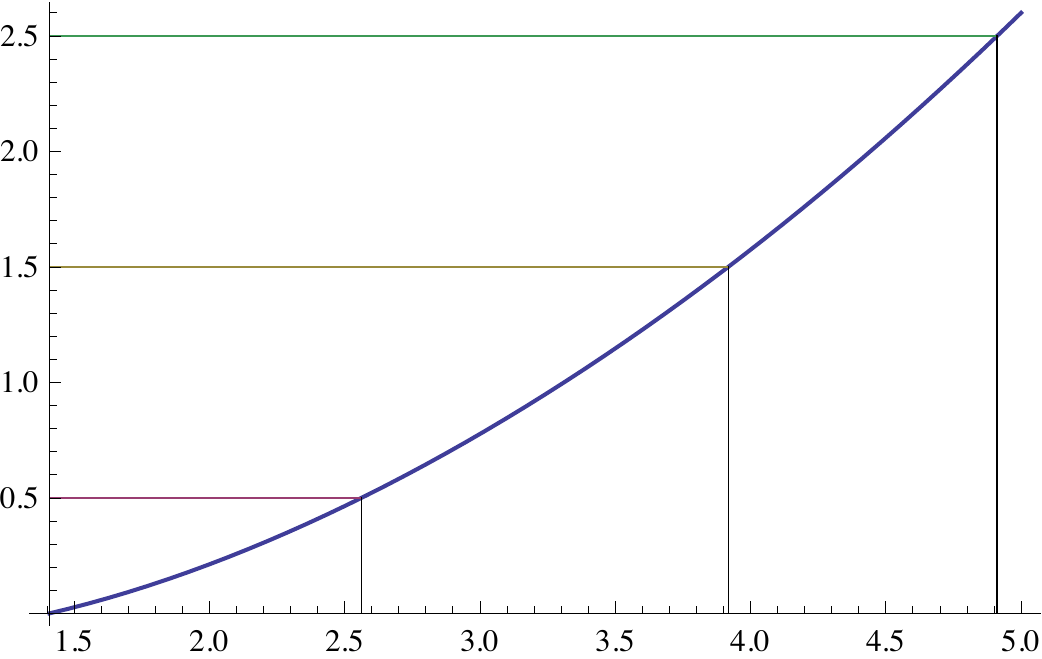}
\caption{The function $\xi(E)-1/4$, as a function of $E$, determining the energy spectrum of the operator (\ref{op-p2}) for $\hbar=2 \pi$.
}
\label{pots}
\end{figure}

It turns out that in the case $\hbar=2\pi$, the exact quantization condition is remarkably simple. Let us consider the Picard--Fuchs equation, 
\be
\label{pf-p2}
\left( \theta^3 -3 z (3 \theta+2) (3 \theta+1)\theta \right) \Pi=0, 
\ee
where 
\be
\theta= z{\rd \over \rd z}.
\ee
This equation has three independent solutions at $z=0$: a trivial, constant solution, a logarithmic solution $\varpi_1(z)$ and a double logarithmic solution $\varpi_2(z)$. If we introduce the power series, 
\be
\label{pers}
\ba
\widetilde \varpi_1(z)&= \sum_{j\ge 1} 3 {(3j-1)! \over (j!)^3} (-z)^j, \\
\widetilde \varpi_2(z)&=\sum_{j \ge 1}{ 18\over j!} 
{  \Gamma( 3j ) 
\over \Gamma (1 + j)^2} \left\{ \psi(3j) - \psi (j+1)\right\}(-z)^{j}, 
\ea
\ee
where $\psi(z)$ is the digamma function, we have that
\be
\label{pers-2}
\ba
\varpi_1(z)&= \log(z) + \widetilde \varpi_1(z), \\
\varpi_2(z)&=\log^2(z) + 2  \widetilde \varpi_1(z) \log(z) +  \widetilde \varpi_2(z). 
\ea
\ee
It is easy to see that the series in (\ref{pers}) have a radius of convergence $|z|=1/27$. The solutions $\varpi_{1,2}(z)$ can be written in terms of 
hypergeometric and Meijer functions. Let us now 
introduce the following function, 
\be
\label{xie}
\xi(E)= {1\over 8 \pi^2} {\varpi_1(z) \varpi_2'(z) - \varpi_2(z) \varpi_1'(z) \over  \varpi_1'(z)}, 
\ee
where
\be
z=\re^{-3 E}. 
\ee
Then, the conjecture of \cite{GHM} implies that the spectrum of (\ref{op-p2}), for $\hbar=2\pi$, is determined by the condition 
\be
\label{p2qc} \xi(E)-{1\over 4} =n+{1\over 2}, \qquad n=0, 1, \cdots. 
\ee
It can be easily seen that the values of $E_n$ determined in this way agree with the ones in Table \ref{table-p2}. A graphical representation of the function $\xi(E)-1/4$, showing the 
different energy levels, can be found in Fig. \ref{pots}. 

What is the content of the quantization condition (\ref{p2qc})? The function $\xi(E)$ has a convergent power series expansion of the form, 
\be
\xi(E)= {9 E^2 \over 8 \pi^2}+ E \sum_{n=1}^\infty a_n \re^{-3 n E}+  \sum_{n=1}^\infty b_n \re^{-3 n E}. 
\ee
When $E$ is large, we can approximate 
\be
\xi(E)-{1\over 4} \approx {9 E^2 \over 8 \pi^2}, 
\ee
and we recover the Bohr--Sommerfeld estimate (\ref{bs}) of the previous section. The constant $1/4$ is a correction which 
can be easily obtained with the next-to-leading WKB method \cite{KaMa,HW,GHM}, and then we have an infinite series of exponentially small corrections in the energy. 
As we will explain later on, they have a mixed physical origin: they combine higher order, perturbative WKB corrections with non-perturbative, instanton-type 
corrections. 

The reader familiar with local mirror symmetry has surely recognized the equation (\ref{pf-p2}): it is the Picard--Fuchs equation governing the periods (\ref{pers-2}) 
of the mirror of local $\IP^2$, which fully determine the genus zero Gromov--Witten invariants of this CY. The variable $z$ is a modulus for the mirror curve, and it is 
related to $\kappa$ in (\ref{mp2}) by 
\be
z= \kappa^{-3}. 
\ee
Therefore, the corrections to the Bohr--Sommerfeld 
quantization condition are governed by the topological string on local $\IP^2$, whose mirror is precisely the algebraic curve (\ref{mp2})! 
As we will see in the next section, this is not a coincidence, but a general story which seems to hold for {\it all} toric CY threefolds: the spectral theory of the operators obtained by 
quantizing their mirror curves is solved by their enumerative invariants.

\section{From topological strings to spectral theory}
\label{st-ts-sec}

\subsection{Mirror symmetry and topological strings} 

In this section we will make a quick summary of the relevant background of mirror symmetry and topological string theory. Useful 
reviews of these subjects might be found in \cite{CK,BMB}. 

Mirror symmetry relates pairs of CY threefolds, $X$, $\widehat X$, in such a way that the enumerative 
geometry of $X$ is reformulated in terms of the deformation of complex structures of $\widehat X$. There are various types of enumerative invariants of $X$. 
Let us consider holomorphic maps from a Riemann surface of genus $g$ to the CY $X$, 
\be
\label{fsigma}
f: \Sigma_g \rightarrow X. 
\ee
Let $[S_i]\in H_2(X,\IZ)$, $i=1, \cdots, s$, be a basis for the two-homology of $X$, with $s=b_2(X)$. The maps (\ref{fsigma}) are classified topologically by the homology class
\be
f_*[(\Sigma_g)]= \sum_{i=1}^s d_i [\Sigma_i] \in H_2(X, \IZ), 
\ee
where $d_i$ are integers called the {\it degrees} of the map. We will put them together in a degree vector ${\bf d}=(d_1, \cdots, d_s)$. 
The {\it Gromov--Witten invariant} at genus $g$ and degree ${\bf d}$, which we will denote by $N_g^{ {\bf d}}$, 
``counts" (in an appropriate way) the number of holomorphic maps of degree ${\bf d}$ from a Riemann 
surface of genus $g$ to the CY $X$. Due to the nature of the moduli space of maps, 
these invariants are in general rational, rather than integer, numbers; see for example \cite{CK} for rigorous definitions and examples. 

The Gromov--Witten invariants at fixed genus $g$ but at all degrees can be put together in generating functionals $F_g({\bf t})$, usually 
called genus $g$ free energies. These are formal power series in $\re^{-t_i}$, $i=1, \cdots, s$, 
where $t_i$ are the {\it K\"ahler parameters} of $X$. More precisely, 
the $t_i$ are flat coordinates on the moduli space of K\"ahler structures of $X$. It is convenient to add to these generating functionals polynomial terms which appear naturally in 
the study of mirror symmetry and topological strings. In this way, we have, at genus zero, 
\be
\label{gzp}
F_0({\bf t})={1\over 6} \sum_{i,j,k=1}^s a_{ijk} t_i t_j t_k  + \sum_{{\bf d}} N_0^{ {\bf d}} \re^{-{\bf d} \cdot {\bf t}}. 
\ee
In the case of a compact CY threefold, the numbers $a_{ijk}$ are interpreted as triple intersection numbers of two-classes in $X$. At genus one, one has
\be
\label{gop}
F_1({\bf t})=\sum_{i=1}^s b_i t_i + \sum_{{\bf d}} N_1^{ {\bf d}} \re^{-{\bf d} \cdot {\bf t}}. 
\ee
In the compact case, the coefficients $b_i$ are related to the second Chern class of the CY manifold \cite{bcov-1}. At higher genus one finds 
\be
\label{genus-g}
F_g({\bf t})= C_g+\sum_{{\bf d}} N_g^{ {\bf d}} \re^{-{\bf d} \cdot {\bf t}}, \qquad g\ge 2, 
\ee
where $C_g$ is a constant, called the constant map contribution to the free energy \cite{BCOV}. 
It turns out that these functionals have a physical interpretation as the free energies at genus $g$ of the so-called type A topological string on $X$. Roughly speaking, 
this free energy can be computed by considering the path integral of a string theory on Riemann surfaces or ``worldsheets" of genus $g$. This makes it possible to use 
a large amount of methods and ideas of physics in order to shed light on these quantities. 

Although the above generating functionals are in principle formal generating functions, they have a common region of convergence near 
the large radius point $t_i \rightarrow \infty$. The total free energy of the topological string is formally defined as the sum, 
\be
\label{tfe}
F^{\rm WS}\left({\bf t}, g_s\right)= \sum_{g\ge 0} g_s^{2g-2} F_g({\bf t})=F^{({\rm p})}({\bf t}, g_s)+ \sum_{g\ge 0} \sum_{\bf d} N_g^{ {\bf d}} \re^{-{\bf d} \cdot {\bf t}} g_s^{2g-2},   
\ee
where
\be
F^{({\rm p})}({\bf t}, g_s)= {1\over 6 g_s^2} \sum_{i,j,k=1}^s a_{ijk} t_i t_j t_k + \sum_{i=1}^s b_i t_i + \sum_{g \ge 2}  C_g g_s^{2g-2}. 
\ee
The superscript ``WS" refers to worldsheet instantons, which are counted by this generating functional. The variable $g_s$, called the {\it topological string coupling constant}, 
is in principle a formal variable keeping track of the genus. However, in string theory this constant has a physical meaning, and measures the 
strength of the string interaction. When $g_s$ is very small, only Riemann surfaces of low genus contribute to a given quantum observable. On the contrary, if $g_s$ is large, the contribution of higher genus 
Riemann surfaces becomes very important. 

The convergence properties of the formal series (\ref{tfe}) are less understood, but there is strong evidence that, for fixed $t_i$ in the common region of convergence, the numerical 
series $F_g({\bf t})$ diverges factorially, as $(2g)!$ (see \cite{mm-nprev} and references therein.) Therefore, the total free energy (\ref{tfe}) 
does {\it not} define in principle a function of $g_s$ and $t_i$. 
In a remarkable paper \cite{GV}, Gopakumar and Vafa 
pointed out that the series (\ref{tfe}) 
can be however resummed order by order in $\exp(-t_i)$, at all orders in $g_s$. This resummation involves a new 
set of enumerative invariants, the so-called {\it Gopakumar--Vafa 
invariants} $n^{\bf d}_g$. Out of these invariants, one constructs the generating series 
\be
\label{GVgf}
F^{\rm GV}\left({\bf t}, g_s\right)=\sum_{g\ge 0} \sum_{\bf d} \sum_{w=1}^\infty {1\over w} n_g^{ {\bf d}} \left(2 \sin { w g_s \over 2} \right)^{2g-2} \re^{-w {\bf d} \cdot {\bf t}}, 
\ee
and one has, as an equality of formal series, 
\be
\label{gv-form}
F^{\rm WS}\left({\bf t}, g_s\right)=F^{({\rm p})}({\bf t}, g_s)+F^{\rm GV}\left({\bf t}, g_s\right). \ee
The Gopakumar--Vafa invariants turn out to be integers, in contrast to the original Gromov--Witten invariants. One can obtain one set of invariants from the other by comparing 
(\ref{tfe}) to (\ref{gv-form}), but there exist direct mathematical constructions of the Gopakumar--Vafa invariants as well, see \cite{PT}. 

The mirror manifold $\widehat X$ to $X$ has $s$ complex deformation parameters $z_i$, $i=1,\cdots, s$, which are related to 
the K\"ahler parameters of $X$ by the so-called {\it mirror map} $t_i( {\boldsymbol z})$. At genus zero, the theory of deformation of 
complex structures of $\widehat X$ can be formulated 
as a theory of periods for the holomorphic $(3,0)$ form of the CY $\widehat X$, which vary 
with the complex structure moduli $z_i$. As first noted in \cite{CGPO}, this theory is remarkably simple. In particular, one can determine 
a function $F_0({\boldsymbol z})$ depending on the complex moduli, called in this context the {\it prepotential}, which is the generating functional of 
genus zero Gromov--Witten invariants (\ref{gzp}), once the mirror map is used. The theory of deformation of complex structures also 
has a physical realization 
in terms of the so-called type B topological string.  
However, for general CY manifolds, a precise definition of this theory at higher genus is still lacking. 
In practice, one often uses the description of the B-model provided in \cite{BCOV}. This 
provides a set of constraints for a series of functions $F_g ({\boldsymbol z})$, $g\ge 1$, known as 
holomorphic anomaly equations. After using the mirror map, these functions become 
the generating functionals (\ref{genus-g}). 

One of the consequences of mirror symmetry and the B-model is the existence of many possible descriptions of the topological string, 
related by symplectic transformations. The existence of these descriptions can be regarded as a generalization of electric-magnetic duality. In this way, 
one finds different ``frames" in which the topological string amplitudes can be expressed. Although these frames are in principle equivalent, some of them might be more 
convenient, depending on the region of moduli space we are looking at. The topological string amplitudes $F_g({\bf t})$ written down above, in terms of Gromov--Witten invariants, 
correspond to the so-called large radius frame in the B-model, and they are appropriate for the so-called large radius limit ${\rm Re}(t_i) \gg 1$. The topological 
string free energies in different frames are related by a formal Fourier--Laplace transform, as explained in \cite{abk}. 

Although mirror symmetry was formulated originally for compact CY threefolds, 
one can extend it to the so-called ``local" case \cite{KKV,CKYZ}. In local mirror symmetry, the CY $X$ is taken to be 
a toric CY manifold, which is necessarily non-compact. The theory of deformation of complex structures of the mirror $\widehat X$ is encoded in an algebraic curve of the form
\be
\label{Wxy}
W(\re^x, \re^y)=0,  
\ee
where the variables appear naturally exponentiated. Local mirror symmetry is considerably simpler than full-fledged mirror symmetry. On the A-model side, 
the enumerative invariants can be computed algorithmically in various ways, either by localization \cite{Ko,GP,KZ}, or 
by using the so-called topological vertex \cite{AKMV}. 
At the same time, the theory of deformation of complex structures of $\widehat X$ can 
be simplified very much. At genus zero, one should consider the periods of the differential
\be
\label{diff-la}
\lambda=y(x) \rd x
\ee
on the curve (\ref{Wxy}) \cite{KKV,CKYZ}. The mirror map and the genus zero free energy $F_0 ({\bf t})$ in the large radius frame are determined by making an 
appropriate choice of cycles on the curve, $\alpha_i$, $\beta_i$, $i=1, \cdots,s$, and one finds
\be
\label{mir-sym}
t_i = \oint_{\alpha_i} \lambda, \qquad \qquad  {\partial F_0 \over \partial t_i} = \oint_{\beta_i} \lambda, \qquad i=1, \cdots, s. 
\ee
In general, $s\ge g_\Sigma$, where $g_\Sigma$ is the genus of the mirror curve. Also, in the local case, the type B topological string can be 
formulated in a more precise way, by using the topological recursion of \cite{eo1}, 
in terms of periods and residues of meromorphic forms on the curve (\ref{Wxy}) \cite{mmopen,bkmp}. 
In particular, mirror symmetry can be proved to 
all genera \cite{eo,flz} by comparing the definition of the B model in \cite{mmopen,bkmp} with localization computations in the A model. 

\begin{example} {\it Local $\IP^2$}. In the case of the local $\IP^2$ CY given in (\ref{Xlocalp2}), the genus zero free energy can be obtained by mirror symmetry as follows. 
The complex deformation parameter in the mirror curve is the parameter $z$ appearing in (\ref{mp2}). The mirror map is given by 
\be
\label{lp2-mm}
t=-\varpi_1(z), 
\ee
where $\varpi_1(z)$ is the power series in (\ref{pers-2}). Then, $F_0(t)$ is defined, up to a constant, by 
\be
\label{f0-def}
{\partial F_0 \over \partial t}= {\varpi_2(z) \over 6}, 
\ee
where $\varpi_2(z)$ is the other period in (\ref{pers-2}). One then finds, after fixing the integration constant appropriately, 
\be
\label{f0-ex}
F_0(t)={t^3 \over 18} + 3 \re^{-t} - {45 \over 8} \re^{-2t} + {244 \over 9} \re^{-3t}+ \cdots. 
\ee
From here one can read the very first Gromov--Witten invariants, like for example $N_0^{d=1}=3$, see \cite{CKYZ} for more details. \qed
\end{example}

Another interesting feature of the local case is that it is possible to define a more general set of enumerative invariants, and consequently a more general 
topological string theory, known sometimes as the {\it refined} topological string. This refinement  has its roots in the instanton partition functions of Nekrasov for supersymmetric 
gauge theories \cite{N}. 
Different aspects of the refinement have been worked out in for example \cite{IKV,KW,HK}. In particular, one can generalize the 
Gopakumar--Vafa invariants to the so-called {\it refined BPS invariants}. Precise mathematical definitions can be found in \cite{CKK,NO}. 
These invariants, which are also integers, depend on the degrees ${\bf d}$ and on two non-negative 
half-integers, $j_L$, $j_R$, or ``spins". We will denote them by $N^{\bf d}_{j_L, j_R}$. The Gopakumar--Vafa invariants are particular combinations 
of these refined BPS invariants, and one has the following relationship, 
\be
\label{ref-gv}
\sum_{j_L, j_R} \chi_{j_L}(q) (2j_R+1) N^{\bf d} _{j_L, j_R} = \sum_{g\ge 0} n_g^{\bf d} \left(q^{1/2}- q^{-1/2} \right)^{2g}, 
\ee
where $q$ is a formal variable and 
\be
\chi_{j}(q)= {q^{2j+1}- q^{-2j-1} \over q-q^{-1}}
\ee
is the $SU(2)$ character for the spin $j$. We note that the sums in (\ref{ref-gv}) are well-defined, since for given degrees ${\bf d}$ only a finite number of $j_L$, $j_R$, $g$ give a non-zero 
contribution. 

Out of these refined BPS invariants, one can define the Nekrasov--Shatashvili (NS) free energy, 
\be
\label{NS-j}
\ba
F^{\rm NS}({\bf t}, \hbar) &={1\over 6 \hbar} \sum_{i,j,k=1}^s a_{ijk} t_i t_j t_k +\sum_{i=1}^s b^{\rm NS}_i t_i \hbar \\
& +\sum_{j_L, j_R} \sum_{w, {\bf d} } 
N^{{\bf d}}_{j_L, j_R}  \frac{\sin\frac{\hbar w}{2}(2j_L+1)\sin\frac{\hbar w}{2}(2j_R+1)}{2 w^2 \sin^3\frac{\hbar w}{2}} \re^{-w {\bf d}\cdot{\bf  t}}. 
\ea
\ee
In this equation, the coefficients $a_{ijk}$ are the same ones that appear in (\ref{gzp}), 
while $b_i^{\rm NS}$ can be obtained by using mirror symmetry as in \cite{KW,HK}. 
The free energy (\ref{NS-j}) is not the most general generating functional for the refined BPS invariants, but 
involves a particular combination thereof, 
which defines the so-called {\it NS limit} of the refined topological string. This limit was first discussed in the 
context of gauge theory in \cite{NS}. By expanding (\ref{NS-j}) in powers of $\hbar$, we find the NS free energies at order $n$, $F^{\rm NS}_n ({\bf t})$, as
\be
\label{ns-expansion}
F^{\rm NS}({\bf t}, \hbar)=\sum_{n=0}^\infty  F^{\rm NS}_n ({\bf t}) \hbar^{2n-1}.
\ee
The expression (\ref{NS-j}) can be regarded as a Gopakumar--Vafa-like resummation of the series in (\ref{ns-expansion}). An important observation is that the first term 
in this series, $F_0^{\rm NS}({\bf t})$, is equal to $F_0({\bf t})$, the standard genus zero free energy. Note that the term involving the coefficients $b_i^{\rm NS}$ contributes to 
$F_1^{\rm NS}({\bf t})$.

\subsection{Quantizing curves}

The semiclassical limit of many interesting problems in quantum physics turns out to be described by a plane curve, together with a choice of a 
meromorphic differential on it. Perhaps the simplest example is the Bohr--Sommerfeld quantization condition (\ref{BS}) for the one-dimensional Schr\"odinger equation. 
A more complicated example is the planar limit of random matrix models, which is typically encoded in a spectral curve which captures the 
information on the limiting distribution of eigenvalues (here, the semiclassical limit is the large $N$ limit). Another example which has received much attention 
recently is the semiclassical limit of ${\rm SL}(2, \IC)$ Chern--Simons theory on the complement of a knot. In this case, the relevant algebraic 
curve is the A-polynomial of a knot \cite{G}. 

The genus zero free energy of topological string theory can be formally regarded as a semiclassical limit, 
in which the r\^ole of $\hbar$ is played by the string coupling constant $g_s$: in the limit $g_s\rightarrow 0$, the dominant 
term in the genus expansion of (\ref{tfe}) is indeed the genus zero free energy. As we explained in (\ref{mir-sym}), this semiclassical limit can be obtained by 
computing periods of the meromorphic differential (\ref{diff}) on the mirror curve.

 It is tempting to think that, in a situation where an algebraic curve encodes the classical 
limit of a quantum problem, the quantum corrections are obtained by considering a ``quantum" version of the curve. This 
has led to a rich literature on ``quantum curves," where the original curve is typically promoted to a 
differential operator (see for example \cite{No,gs} for recent reviews). In the case of topological string theory, 
an approach based on quantum curves was first proposed in \cite{ADKMV}, where it was suggested that the higher genus free energies 
of topological string theory (which can be regarded as ``quantum corrections") might be obtained by quantizing the mirror curve in an appropriate way. Building on work 
on supersymmetric gauge theories and quantum integrable systems \cite{NS}, it was later shown in \cite{ACDKV} that mirror curves can be 
formally quantized by using the WKB approximation (this was first pointed out in \cite{MM}, in the context of Seiberg--Witten curves). 
However, the quantum corrections obtained in this way do not give the conventional higher genus Gromov--Witten invariants, 
but rather the NS free energy (\ref{ns-expansion}). 

It is well-known that the quantization of a classical system is in general not unique: 
there can be different quantizations of the same function, 
leading to the same classical limit but differing when $\hbar\not=0$.
The quantization of an algebraic curve is also plagued with ambiguities, for at least three different reasons. 
First of all, the quantization of a curve inherits the 
ordering ambiguities of any quantization procedure. Second, algebraic curves are naturally defined in the complex domain, 
while their quantum versions typically involve a choice of reality conditions 
which can be made in different ways. Finally, when an algebraic curve is 
promoted to an operator, one has to specify the Hilbert space of wavefunctions where this operator acts, 
and in particular the boundary conditions satisfied by 
such wavefunctions. In most of the recent literature on ``quantum curves," these issues are not addressed explicitly, 
and quantum curves are only studied at the level of formal WKB expansions, as in \cite{ACDKV}. 
This perturbative approach has led to many interesting results, 
but does not give a non-perturbative formulation of the quantum problem.  

In \cite{KaMa} it was pointed out that the mirror curves of toric CY threefolds can be quantized 
in such a way that the resulting operators have a discrete spectrum (we have already 
seen an example in section \ref{problem}, in the case of local $\IP^2$.) It was then conjectured in \cite{GHM} that {\it the quantization of mirror curves leads to 
positive definite, trace class operators on $L^2(\IR)$}. 
This was proved in \cite{KM} in many examples. In the scheme proposed in \cite{GHM,KM}, the ambiguities arising 
in the quantization of algebraic curves are solved in a natural way: 
ordering ambiguities are fixed by using Weyl quantization, 
which is particularly well suited for exponentiated position and momentum operators. Reality conditions are chosen in such a way that the 
classical regions (\ref{re}) in phase space are compact. Finally, the Hilbert space where the operators act is simply $L^2(\IR)$. 

In the rest of this paper, we will focus for simplicity on toric (almost) del Pezzo CY threefolds. These CYs are 
defined as the total space of the canonical bundle on a toric (almost) del Pezzo surface $S$,
\be
\label{dP}
X=\CO(K_S) \rightarrow S. 
\ee
They are sometimes called ``local $S$." For example, if $S= \IP^2$, the total space of its canonical bundle will be called local $\IP^2$. Examples 
of toric del Pezzos include, besides $\IP^2$, the  Hirzebruch surfaces $\IF_n$, $n=0, 1,2$, and the blowups of $\IP^2$ at $n$ points, denoted by 
${\mathcal B}_n$, for $n=1,2,3$ (note that $\IF_1= {\mathcal B}_1$, and that $\IF_0= \IP^1 \times \IP^1$). The main simplifying characteristic of these manifolds is that 
their mirror curve has genus one. This makes their analysis much simpler. We hasten to add that there is a very interesting generalization 
of all the considerations in this review paper to mirror curves of higher genus \cite{cgm2}. In this case, a mirror curve of genus $g_\Sigma$ leads in general to $g_\Sigma$ trace class operators. 

\begin{figure}[h]
\center
\includegraphics[scale=0.4]{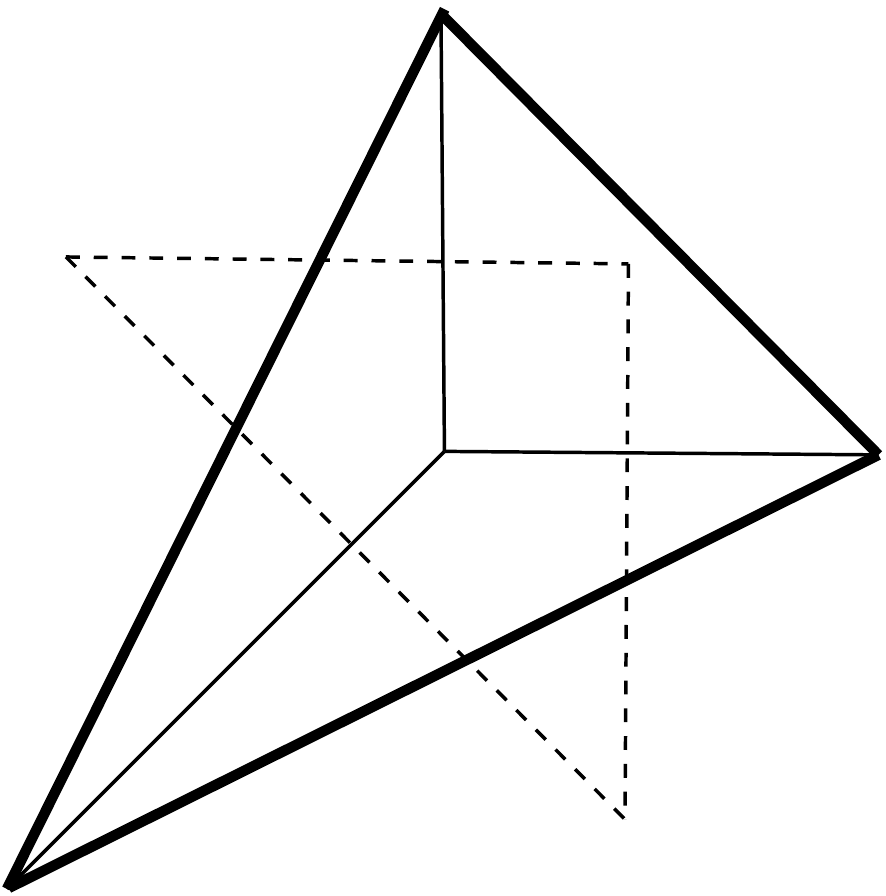}
\caption{The vectors (\ref{p2-vec}) defining the local $\IP^2$ geometry, together with the polyhedron $\Delta_{\IP^2}$ (in thick lines) and the 
dual polyhedron (in dashed lines).
}
\label{hut}
\end{figure}

By standard results in toric geometry (see for example \cite{HKP,CR}), 
toric, almost del Pezzo surfaces can be classified by reflexive polyhedra in two dimensions. The polyhedron $\Delta_S$ associated to a surface $S$ is 
the convex hull of a set of two-dimensional vectors 
\be
\label{vector}
\nu^{(i)}=\left(\nu^{(i)}_1, \nu^{(i)}_2\right), \qquad i=1, \cdots, s+2, 
\ee
together with the origin, see \figref{hut} for the example of local $\IP^2$. In order to construct the total space of the canonical bundle over $S$, we have to consider 
the extended vectors 
\be
\ba
\overline \nu^{(0)}&=(1, 0,0), \\
\overline \nu^{(i)}&=\left(1, \nu^{(i)}_1, \nu^{(i)}_2\right),\qquad i=1, \cdots, s+2. 
\ea
\ee
They satisfy the relations 
\be
\sum_{i=0}^{s+2} Q^\alpha_i \overline \nu^{(i)}=0, \qquad \alpha=1, \cdots, s, 
\ee
where $Q^\alpha_i$ is a matrix of integers (called the charge matrix) which characterizes the geometry. 

The construction of the mirror geometry to (\ref{dP}) goes back to Batyrev, and it has been recently reviewed in \cite{CR}, 
to which we refer for further details. 
In order to write down the equation for the mirror curve of (\ref{dP}), we note that the $s$ complex parameters of the mirror can be divided in 
two types: one ``true" modulus $\kappa$ (for 
a curve of genus one) and a set of ``mass" parameters $\xi_i$, $i=1, \cdots, s-1$ \cite{HKP,HKRS}. In terms of 
these variables, the mirror curve for a local del Pezzo CY threefold can be written as, 
\be
\label{ex-W}
W(\re^x, \re^y)= \CO_S(x,y)+ \kappa=0,  
\ee
where
\be
\label{coxp}
 \CO_S (x,y)=\sum_{i=1}^{s+2} \exp\left( \nu^{(i)}_1 x+  \nu^{(i)}_2 y + f_i(\boldsymbol{\xi}) \right), 
 \ee
and $f_i(\boldsymbol{\xi})$ are suitable functions of the parameters $\xi_j$. 

\begin{example} The simplest case of a local del Pezzo CY is local $\IP^2$. In this case, we have $s=1$. 
The vectors (\ref{vector}) are given by 
\be
\label{p2-vec}
\nu^{(1)}=(1,0), \qquad \nu^{(2)}=(0,1), \qquad \nu^{(3)}=(-1,-1). 
\ee
We show these vectors, together with the origin and their convex hull $\Delta_{\IP^2}$, in \figref{hut}. In this geometry there is one complex 
deformation parameter $\kappa$, and the function 
$\CO_{\IP^2}(x,y)$ is given by  
\be
\label{lp2}
\CO_{\IP^2} \left(x, y \right)= \re^{ x} + \re^{y} + \re^{- x- y}. 
\ee
This is the curve (\ref{mp2}). Note that, as shown in \figref{hut}, the dual polyhedron to $\Delta_{\IP^2}$ is precisely the region in \figref{p2-reg}.  \qed
\end{example}
 
\begin{example} 
The previous example can be generalized by considering the canonical bundle over the 
weighted projective space $\IP (1, m, n)$, where $m, n \in \BZ_{>0}$. This is not a smooth manifold, but 
it can be analyzed by using extensions of Gromov--Witten theory, see for example \cite{BC} for a study of the case $n=1$. The vectors are in this case 
\be
\nu^{(1)}=(1,0), \qquad \nu^{(2)}=(0,1), \qquad \nu^{(3)}=(-m,-n), 
\ee
and the function appearing in the mirror curve (\ref{ex-W}) is given by 
\be
\label{lpnm}
\CO_{m,n} \left(x, y \right)= \re^{ x} + \re^{y} + \re^{- m x- n y}. 
\ee
Some of these geometries can arise as degeneration limits of toric del Pezzos. For example, the 
mirror curve to local $\IF_2$ is characterized by the function 
\be
\CO_{\IF_2} \left(x, y \right)= \re^{ x} + \re^{y} + \re^{- 2 x- y} + \xi \re^{-x}, 
\ee
and when $\xi=0$ we recover the geometry (\ref{lpnm}) with $m=2$ and $n=1$. \qed
\end{example} 

Some further examples of functions  obtained by quantization of local del Pezzos can be found in Table \ref{table-ops}. Details on the corresponding 
geometries can be found in for example \cite{HKP}. 

\begin{table}
\centering
\begin{tabular}{||  l || l || l || l ||}
\hline
$S$  & $\CO_S (x,y)$  \\ \hline\hline
$\IP^2$ &  $\re^{x}+ \re^y + \re^{-x-y}$  \\ \hline \hline
$\IF_0$ &  $\re^x+ \xi \re^{-x} + \re^y + \re^{-y}$  \\ \hline \hline
$\IF_1$ &  $\re^x+ \re^y +  \re^{-x -y}+\xi \re^{-x} $  \\ \hline \hline
$\IF_2$ & $\re^x+  \re^y + \re^{ -2x-y }+\xi \re^{-x}$ \\ \hline \hline
$\CB_2$ & $ \re^{x} +  \re^{y } +  \re^{-x-y} +\xi_1 \re^{-y}+ \xi_2 \re^{-x}$  \\ \hline \hline
$\CB_3$ & $ \re^x +  \re^y+ \re^{-x-y} + \xi_1 \re^{-x} +\xi_2 \re^{-y}+ \xi_3 \re^{x+ y}$ \\ \hline 
\end{tabular}
\vskip .5cm
\caption{The functions $\CO_S(x,y)$ associated to some local del Pezzo CYs.}
\label{table-ops}
\end{table}

The ``quantization" of the mirror curve (\ref{ex-W}), in the case of local del Pezzo CYs, is based 
on the promotion of the function $\CO_S(x,y)$ to an operator, which 
will be denoted by $\mathsf{O}_S$. First, we promote $x$, $y$ to self-adjoint Heisenberg 
operators $\mathsf{x}$, $\mathsf{y}$ on the real line, satisfying the commutation relation (\ref{hcr}). This involves a choice of reality conditions for the complex variables 
$x$, $y$. Then, we apply Weyl's quantization to $\CO_S(x,y)$. In particular, this fixes possible ordering ambiguities. We recall that, in Weyl's quantization, we have 
\be
\re^{r x + s y} \rightarrow \re^{r \mx + s \my}. 
\ee
As noted in \cite{GHM}, instead of focusing on the operator $\mathsf{O}_S$ (which is not of trace class), one should rather consider its inverse 
\be
\label{rhos}
\rho_S=\mathsf{O}^{-1}_S
\ee
acting on $L^2(\IR)$. It was conjectured in \cite{GHM} that the operators $\rho_S$ are 
of trace class and positive definite, provided appropriate positivity conditions are imposed on the mass parameters appearing in the mirror curves. 

In order to motivate this assertion, note that, if $\re^{E_n}$ are the eigenvalues of $\mO_S$, the eigenvalues of $\rho_S$ are 
$\re^{-E_n}$. It is easy to extend the Bohr--Sommerfeld 
estimate (\ref{bs}) to all the operators $\mO_S$ (see \cite{GHM}): we can obviously define an available region in phase space, $\CR(E)$, similar to what we did in (\ref{re}), but involving this time 
the classical curve $\CO_S$. The volume of this region behaves, at large $E$, as
\be
\label{large-vol}
{\rm vol}_0(E) \approx C E^2, \qquad E\gg 1. 
\ee
The coefficient $C$ can be easily computed by considering the tropical limit of the classical curve, where the region $\CR(E)$ becomes a polygon (in fact, it is the dual polyhedron to $\Delta_S$). We then find the 
estimate 
\be
\label{gen-estimate}
E_n \approx {\sqrt{2 \pi \hbar \over C}} n^{1/2}, \qquad n \gg 1. 
\ee
This heuristic argument indicates that the {\it spectral traces} of the operators $\rho_S$, 
\be
\tr \rho_{S}^\ell = \sum_{n\ge 0} \re^{-\ell E_n}, \qquad \ell=1, 2, \cdots
\ee
are finite. We then expect $\rho_S$ to be positive-definite (at least for some range of the parameters) and of trace class. This was proved in \cite{KM} for all the operators 
appearing in Table \ref{table-ops}. Moreover, in some cases it is possible to calculate the exact integral kernel of the corresponding trace class operator \cite{KM,KMZ,CGuM}. 
An important example in this respect is the family of operators associated to the function (\ref{lpnm}):
 \begin{equation}
 \label{tto}
\mathsf{O}_{m,n}=\re^{\mathsf{x}}+\re^{\mathsf{y}} +\re^{-m\mathsf{x}-n\mathsf{y}},\quad m,n\in\mathbb{R}_{>0}.
\end{equation}
These operators were called three-term operators in \cite{KM}. In this discussion, $m, n$ are positive, real numbers, although in applications to the quantization 
of mirror curves they are often positive integers (like in the quantization of the mirror curves to local $\IP(1,n,m)$). Let us first introduce some notation. 
As in \cite{KM}, we will denote by $ \fad(x)$ Faddeev's quantum dilogarithm \cite{F,FK}. We define as well 
\be
\label{mypsi-def}
\mypsi{a}{c}(x)= \frac{\re^{2\pi ax}}{\fad(x-\im(a+c))}. 
\ee
We then have the following proposition, proved in \cite{KM}.

\begin{proposition}
The operator on $L^2(\IR)$
\be
\label{rhomn}
\rho_{m,n}=\mO_{m,n}^{-1}
\ee
 is positive-definite and of trace class. Let us define normalized Heisenberg operators $\mq$, $\map$, satisfying the normalized 
commutation relation
\be
[\map, \mq]=(2 \pi \im)^{-1}.
\ee
They are related to $\mx$, $\my$ by the linear canonical transformation, 
\begin{equation}
\mathsf{x}\equiv 2\pi\mathsf{b}\frac{(n+1)\mathsf{p}+n\mathsf{q}}{m+n+1},\quad \mathsf{y}\equiv -2\pi\mathsf{b}\frac{m\mathsf{p}+(m+1)\mathsf{q}}{m+n+1}, 
\end{equation}
so that $\hbar$ is related to $\mb$ by
\be
\label{b-hbar}
\hbar=\frac{2\pi\mathsf{b}^2}{m+n+1}. 
\ee
In the momentum representation associated to $\map$, 
the operator $\rho_{m,n}$ has the integral kernel, 
\begin{equation}
\label{ex-k}
\rho_{m,n}(p,p')=\frac{\overline{\mypsi{a}{c}(p)}\mypsi{a}{c}(p')}{2\mathsf{b}\cosh\left(\pi\frac{p-p'+\im (a+c-nc)}{\mathsf{b}}\right)}.
\end{equation}
In this equation, $a$, $c$ are given by 
\be
a =\frac{m \mb}{2(m+n+1)}, \qquad c=\frac{\mb}{2(m+n+1)}. 
\ee
\end{proposition}

Once the trace class property has been established for the operators $\rho_{m,n}$, it can be easily established for 
operators $\rho_S$ whose inverse 
$\mO_S$ is a perturbation of a three-term operator $\mO_{m,n}$ by a 
positive self-adjoint operator \cite{KM}. In addition, it can be shown that the operator $\rho_S$ for $S= \IF_0$ is also of trace class. This proves the trace class property 
for a large number of operators arising from the quantization of mirror curves, including all the operators appearing in Table \ref{table-ops}. The trace class property of $\rho_{m,n}$ and of $\rho_{\IF_0}$ follows as well 
from the estimate (\ref{gen-estimate}), which was established rigorously in these cases in \cite{LST}. 

\subsection{Fredholm determinants from topological strings} In the previous sections, we have shown that the operators $\rho_S= \mathsf{O}_S^{-1}$ arising from mirror curves, for many 
toric del Pezzo threefolds, are of trace class. This means in particular that their {\it spectral} or {\it Fredholm determinant}, 
\be
\label{f-det}
\Xi_S(\kappa, \hbar)={\rm det} \left( 1+ \kappa \rho_S\right) 
\ee
exists and is an entire function of $\kappa$ \cite{Si}. The Fredholm determinant has as a 
power series expansion around $\kappa=0$, of the form 
\be
\Xi_S(\kappa, \hbar)=1+\sum_{N=1}^\infty Z_S(N, \hbar) \kappa^N, 
\ee
where $Z_S(N, \hbar)$ is the {\it fermionic spectral trace}, given by 
\be
Z_S(N, \hbar) = \tr \left(\Lambda^N(\rho_S)\right),  \qquad N=1,2, \cdots
\ee
In this expression, the operator $\Lambda^N(\rho_S)$ is defined by $\rho_S^{\otimes N}$ acting on $\Lambda^N\left(L^2(\IR)\right)$. 
A theorem of Fredholm asserts that, if $\rho_S(p_i, p_j)$ is 
the kernel of $\rho_S$, the fermionic spectral trace can be computed as a multi-dimensional integral, 
\be
\label{fred}
Z_S(N, \hbar) = {1 \over N!}  \int   {\rm det}\left( \rho_S (p_i, p_j) \right)\, \rd ^N p . 
\ee
In turn, the logarithm of the Fredholm determinant can be regarded as a generating 
functional of the spectral traces of $\rho_S$, since 
\be
\CJ_S (\kappa, \hbar)=\log \Xi_S (\kappa, \hbar)= -\sum_{\ell=1}^\infty {(-\kappa)^\ell \over \ell} \tr \rho_S^\ell.  
\ee
The Fredholm determinant (\ref{f-det}) has the infinite product representation \cite{Si} 
\be
\Xi_S(\kappa, \hbar)=\prod_{n=0}^\infty \left( 1+ \kappa \re^{-E_n} \right), 
\ee
where $\re^{-E_n}$ are the eigenvalues of $\rho_S$. Therefore, one way of obtaining the spectrum of $\rho_S$ is to look for the zeroes of  $\Xi_S$, which occur at
\be
\label{zrs}
\kappa=-\re^{E_n}, \qquad n \ge 0. 
\ee

There are very few operators on $L^2(\IR)$ for which the Fredholm determinant can be written down explicitly. One family of examples which has been studied in 
some detail are Schr\"odinger operators with homogeneous potentials $V(x)=|x|^s$, $s=1, 2, \cdots$ \cite{voros2,DT}. 
In this case, the Fredholm determinant is defined as a regularized version of 
\be 
D(E)=\prod_{n\ge 0}  \left( 1+ {E \over E_n} \right), 
\ee
see for example \cite{voros-sd} for a detailed treatment. For these potentials, the function $D(E)$ is 
known to satisfy certain functional equations and it is captured by integral equations of the TBA type \cite{DT}. There is however no closed formula for it. One of the surprising (conjectural) 
results of \cite{GHM} is that 
the Fredholm determinant of the operators $\rho_S$ can be computed {\it exactly and explicitly} 
in terms of the enumerative invariants of $X$ encoded in the topological string amplitudes. Moreover, 
the zeroes of the Fredholm determinant are determined by {\it exact quantization conditions}. The result stated in (\ref{p2qc}) for the spectrum of $\rho_{\IP^2}$ at $\hbar=2 \pi$ 
is a particular example of the general 
conjecture in \cite{GHM}. The resulting quantization conditions are akin to those found in 
conventional Quantum Mechanics (like in, for example, \cite{ZJJ,voros2}), but with an important difference: they involve {\it convergent} series, and in particular 
one does not need the apparatus of Borel--\'Ecalle resummation in order to determine the spectrum.   

The conjectural expression of \cite{GHM} for the Fredholm determinant of the operator $\rho_S$ requires 
the two generating functionals of enumerative invariants considered before, (\ref{GVgf}) and (\ref{NS-j}). 
In order to state the result, we identify the parameter $\kappa$ appearing in the Fredholm determinant, with the geometric modulus 
of $X$ appearing in (\ref{ex-W}). We will also write $\kappa$ in terms of the ``chemical potential" $\mu$
\be
\kappa =\re^\mu.
\ee
In addition to the modulus $\kappa$, we have the ``mass parameters" $\xi_j$. The flat coordinates for the K\"ahler 
moduli space, $t_i$, are related to $\kappa$ and $\xi_j$ by the mirror map, and at leading order, in the large radius limit, we have 
\be
\label{tmu}
t_i \approx c_i \mu -\sum_{j=1}^{s-1} \alpha_{ij} \log \xi_j,  \quad i=1, \cdots, s, 
\ee
where $c_i$, $\alpha_{ij}$ are constants.  As shown in \cite{ACDKV}, the WKB approach makes it possible to define a 
{\it quantum mirror map} $t_i(\hbar)$, which in the limit $\hbar \rightarrow 0$ agrees with the conventional 
mirror map. It is a function of $\mu$, $\xi_j$ and $\hbar$ and it can be computed as an A-period of a quantum corrected version of the differential (\ref{diff-la}). This 
quantum correction is simply obtained by using the all-orders, perturbative WKB method of Dunham.

We now introduce two different functions of $\mu$. The first one is the WKB grand potential, 
\be
\label{jm2}
\ba
\mathsf{J}^{\rm WKB}_S (\mu, \boldsymbol{\xi}, \hbar)&=\sum_{i=1}^s  {t_i(\hbar) \over 2 \pi}   {\partial F^{\rm NS}({\bf t}(\hbar), \hbar) \over \partial t_i} 
+{\hbar^2 \over 2 \pi} {\partial \over \partial \hbar} \left(  {F^{\rm NS}({\bf t}(\hbar), \hbar) \over \hbar} \right)\\
&+  \sum_{i=1}^s  {2 \pi \over \hbar} b_i t_i(\hbar) + A({\boldsymbol \xi}, \hbar). 
\ea
\ee
In this equation, $F^{\rm NS}({\bf t}, \hbar)$ is given by the expression (\ref{NS-j}). In the second term of (\ref{jm2}), the derivative w.r.t. $\hbar$ does not act on the implicit dependence of $t_i$. The coefficients 
$b_i$ appearing in the last line are the same ones appearing in (\ref{gop}). The function $A({\boldsymbol \xi}, \hbar)$ is not 
known in closed form for arbitrary 
geometries, although detailed conjectures for its form exist in various examples. It is closely related to a 
all-genus resummed version of the constant map contribution $C_g$ appearing in (\ref{genus-g}). 
The second function is the ``worldsheet" grand potential, which can be obtained from the generating functional (\ref{GVgf}), 
\be
\label{jws}
\mathsf{J}^{\rm WS}_S(\mu,  \boldsymbol{\xi}, \hbar)= F^{\rm GV}\left( {2 \pi \over \hbar}{\bf t}(\hbar)+ \pi \ri {\bf B}, {4 \pi^2 \over \hbar} \right).
\ee
In this formula, ${\bf B}$ is a constant vector (``B-field") which depends on the geometry under consideration. This vector should satisfy the following requirement: 
for all ${\bf d}$, $j_L$ and $j_R$ such that the refined BPS invariant $N^{{\bf d}}_{j_L, j_R} $ is non-vanishing, we must have
\be
(-1)^{2j_L + 2 j_R+1}= (-1)^{{\bf B} \cdot {\bf d}}. 
\ee
For local del Pezzo CY threefolds, the existence of such a vector was established in \cite{hmmo}. 
Note that the effect of this constant vector is to introduce a sign 
\be
\label{K-sign}
(-1)^{ w {\bf d} \cdot {\bf B}}
\ee
in the generating functional (\ref{GVgf}). An important remark is that, in (\ref{jws}), the topological string coupling constant $g_s$ appearing in (\ref{tfe}) and (\ref{GVgf}) is related to the 
Planck constant appearing in the spectral problem by, 
\be
g_s= {4 \pi^2 \over \hbar}. 
\ee
Therefore, the regime of weak coupling for the topological string coupling constant, $g_s\ll 1$, corresponds to the strong coupling regime of the spectral problem, $\hbar \gg 1$, and conversely, the 
semiclassical limit of the spectral problem corresponds to the strongly coupled topological string. We therefore have a {\it strong-weak coupling duality} between the 
spectral problem and the conventional topological string. 

The {\it total grand potential} is the sum of these two functions, 
\be
\label{jtotal}
\mathsf{J}_S (\mu, \boldsymbol{\xi}, \hbar) = \mathsf{J}^{\rm WKB}_S (\mu, \boldsymbol{\xi}, \hbar)+ \mathsf{J}^{\rm WS}_S (\mu,  \boldsymbol{\xi}, \hbar), 
\ee
and it was first considered in \cite{hmmo}. It has the structure
\be
\mathsf{J}_S (\mu, \boldsymbol{\xi},\hbar)= {1\over 12 \pi \hbar} \sum_{i,j,k=1}^s a_{ijk} t_i t_j t_k + \sum_{i=1}^s \left( {2 \pi b_i \over \hbar} + {\hbar b_i^{\rm NS} \over 2 \pi} \right) t_i + 
\CO\left( \re^{-t_i}, \re^{-2 \pi t_i/\hbar} \right),
\ee
where the last term stands for a formal power series in $\re^{-t_i}$, $\re^{-2 \pi t_i/\hbar}$, whose coefficients depend explicitly 
on $\hbar$. Note that the trigonometric functions appearing in (\ref{NS-j}) and (\ref{GVgf}) have double poles when $\hbar$ is a rational multiple of $\pi$. 
However, as shown in \cite{hmmo}, 
the poles {\it cancel} in the sum (\ref{jtotal}). This HMO cancellation mechanism was first discovered in \cite{HMO}, in a slightly 
different context, and it was first advocated in \cite{KaMa} in the study of quantum curves. 

A natural question is whether the formal power series in (\ref{jtotal}) converges, at least for some values of 
its arguments. Although we do not have rigorous results on this 
problem, the available evidence suggests that, for real $\hbar$, $\mathsf{J}_S(\mu, \boldsymbol{\xi}, \hbar)$ 
{\it converges} in a neighbourhood of the large radius point $t_i\rightarrow \infty$. 
However, the series seems to be divergent when $\hbar$ is complex. This divergence is inherited from the generating 
functionals (\ref{GVgf}) and (\ref{NS-j}). Explicit 
calculations indicate, for example, that the Gopakumar--Vafa 
generating functional (\ref{GVgf}) is divergent for complex $g_s$, as first noted in 
\cite{hmmo}. In the one-modulus case, if we write this series as 
\be
F^{\rm GV}(t, g_s) = \sum_{\ell \ge 1} a_\ell(g_s) \re^{-\ell t}, 
\ee
one finds that, if $g_s \in \IC \backslash \IR$, 
\be
\log \left| a_\ell(g_s) \right|\sim \ell^2, \qquad \ell\gg 1. 
\ee
Therefore, the generating functional (\ref{GVgf}) is in general ill-defined and, as it stands, it can not be used as a non-perturbative definition of the topological string free energy\footnote{In some geometries, (\ref{GVgf}) can be resummed by using instanton calculus, and one obtains a convergent function if $g_s \in \IC \backslash \IR$ \cite{BS}. The resulting function still has a dense set of 
poles on the real line.}. 
Finally, note that, after using the explicit expression for the $t_i$ in terms of $\mu$ and the mass parameters, one easily finds that $\mathsf{J}_S(\mu, \boldsymbol{\xi},\hbar)$ is a 
cubic polynomial in $\mu$, plus an infinite series of exponentially small corrections in $\re^{-\mu}$, $\re^{-2 \pi \mu/\hbar}$. At large $\mu$, the leading behaviour is given by, 
\be
\mathsf{J}_{S}(\mu, \boldsymbol{\xi},\hbar) \approx {C \over 6 \pi \hbar} \mu^3,  \qquad \mu \gg 1, 
\ee
where $C$ is the constant appearing in (\ref{large-vol}). 

We can now state the main conjecture of \cite{GHM}. 

\begin{conjecture} The Fredholm determinant of the operator $\rho_S$ is given by 
\be
\label{spec-det}
\Xi_S (\kappa, \boldsymbol{\xi}, \hbar)= \sum_{n \in \BZ} \exp\left(  \mathsf{J}_S (\mu + 2 \pi \im n, \boldsymbol{\xi}, \hbar) \right).  
\ee
\end{conjecture}

The sum over $n$ defines a {\it quantum theta function} $\Theta_S \left( \mu, \boldsymbol{\xi}, \hbar \right)$, 
\be
\label{gen-theta}
\Xi_S (\kappa, \boldsymbol{\xi}, \hbar)=\re^{ \mathsf{J}_S (\mu, \boldsymbol{\xi}, \hbar) }  \Theta_S \left( \mu, \boldsymbol{\xi}, \hbar \right). 
\ee
The reason for this name is that, when $\hbar=2 \pi$, the quantum theta function becomes a classical, conventional theta function, as we will see in a moment. The vanishing locus 
of the Fredholm determinant, as we have seen in (\ref{zrs}), gives the spectrum of the trace class operator. Given the form of (\ref{gen-theta}), this is the vanishing locus of the 
quantum theta function. 

It would seem that the above conjecture is very difficult to test, since it gives the Fredholm determinant as a formal, infinite sum. However, it is easy to see that the r.h.s. of (\ref{spec-det}) 
has a series expansion at large $\mu$ in powers of $\re^{-\mu}$, $\re^{-2 \pi \mu/\hbar}$. In addition, it leads to an integral representation for the fermionic spectral trace which is very 
useful in practice: if we write $Z_S(N, \hbar)$ as a contour integral around $\kappa=0$, simple manipulations give the expression \cite{HMO,GHM}
\be
\label{int-Z}
Z_{S}(N, \hbar) = {1\over 2 \pi \ri} \int_{\mathcal C} \re^{\mJ_S(\mu,  \boldsymbol{\xi}, \hbar)- N \mu} \rd \mu, 
\ee
where ${\mathcal C}$ is a contour going from $\re^{-\im \pi/3} \infty$ to $\re^{\im \pi/3} \infty$. Note that this is the standard contour for 
the integral representation of the Airy function and it should lead to a convergent integral, since $\mJ_S(\mu,\boldsymbol{\xi}, \hbar)$ is given by a 
cubic polynomial in $\mu$, plus exponentially small corrections. Finally, as we will see in a moment, in some cases the r.h.s. of (\ref{spec-det}) can be written 
in terms of well-defined functions. 

What is the interpretation of the total grand potential that we introduced in (\ref{jtotal})? The 
WKB part takes into account the perturbative corrections (in $\hbar$) to the 
spectral problem defined by the operator $\rho_S$. In fact, it can be calculated order by order in the $\hbar$ expansion, 
by using standard techniques in Quantum and Statistical 
Mechanics (see for example \cite{MP,hatsuda}). The expression (\ref{jm2}) provides a resummation of this 
expansion at large $\mu$. However, the WKB piece is insufficient 
to solve the spectral problem, since as we mentioned above, (\ref{jm2}) is divergent for a dense set of values of $\hbar$ on the real line. 
Physically, the additional generating functional (\ref{jws}) contains the contribution of {\it complex} instantons, which are non-perturbative in $\hbar$ \cite{MP,KaMa} 
and cancel the poles in the all-orders WKB contribution. Surprisingly, 
the full instanton contribution in this spectral problem is simply encoded in the standard topological string partition function. 

The formulae (\ref{jm2}), (\ref{jws}), (\ref{spec-det}) 
are relatively complicated, in the sense that they involve the full generating functionals (\ref{NS-j}) and (\ref{GVgf}). There is however an important case in which 
they simplify considerably, namely, when $\hbar=2 \pi$. This was called in \cite{GHM} the ``maximally supersymmetric case," since in the closely related spectral problem 
of ABJM theory, it occurs when there is enhanced ${\mathcal N}=8$ supersymmetry \cite{CGM}. 
In this case, as it can be easily seen from the explicit expressions for the generating functionals, many contributions vanish. For example, 
in the Gopakumar--Vafa generating functional, all terms involving $g\ge 2$ are zero. A simple calculation shows that the function (\ref{jtotal}) is given by 
\be
\label{msJ}
\mJ_S(\mu, \boldsymbol{\xi}, 2 \pi)=  {1\over 8 \pi^2} \sum_{i, j=1}^s t_i t_j {\partial^2 \widehat F_0 \over \partial t_i \partial t_j} 
-{1\over 4 \pi^2} \sum_{i=1}^s t_i {\partial  \widehat F_0 \over \partial t_i }+{1\over 4 \pi^2}  \widehat F_0  ({\bf t})+  \widehat F_1({\bf t})+  \widehat F_1^{\text{NS}}({\bf t}),  
\ee
where $ \widehat F_0({\bf t})$, $ \widehat F_1({\bf t})$ and $ \widehat F_1^{\text{NS}}({\bf t})$ are the generating 
functions appearing in (\ref{gzp}), (\ref{gop}) and (\ref{ns-expansion}), with the only difference 
that one has to include as well the sign (\ref{K-sign}) in the expansion in $\re^{-t_i}$. It can be also seen that, for this value of 
$\hbar$, the quantum mirror map becomes the classical mirror map, up to a change of sign $\kappa \rightarrow -\kappa$. One can now use (\ref{msJ}) to 
compute the quantum theta function appearing in (\ref{gen-theta}). It is easy to see that, provided some integrality properties hold for the constant $C$ in (\ref{large-vol}), the quantum 
theta function becomes a classical theta function. We will now illustrate this simplification in the case of local $\IP^2$. 

\begin{figure}[h]
\center
\includegraphics[scale=0.55]{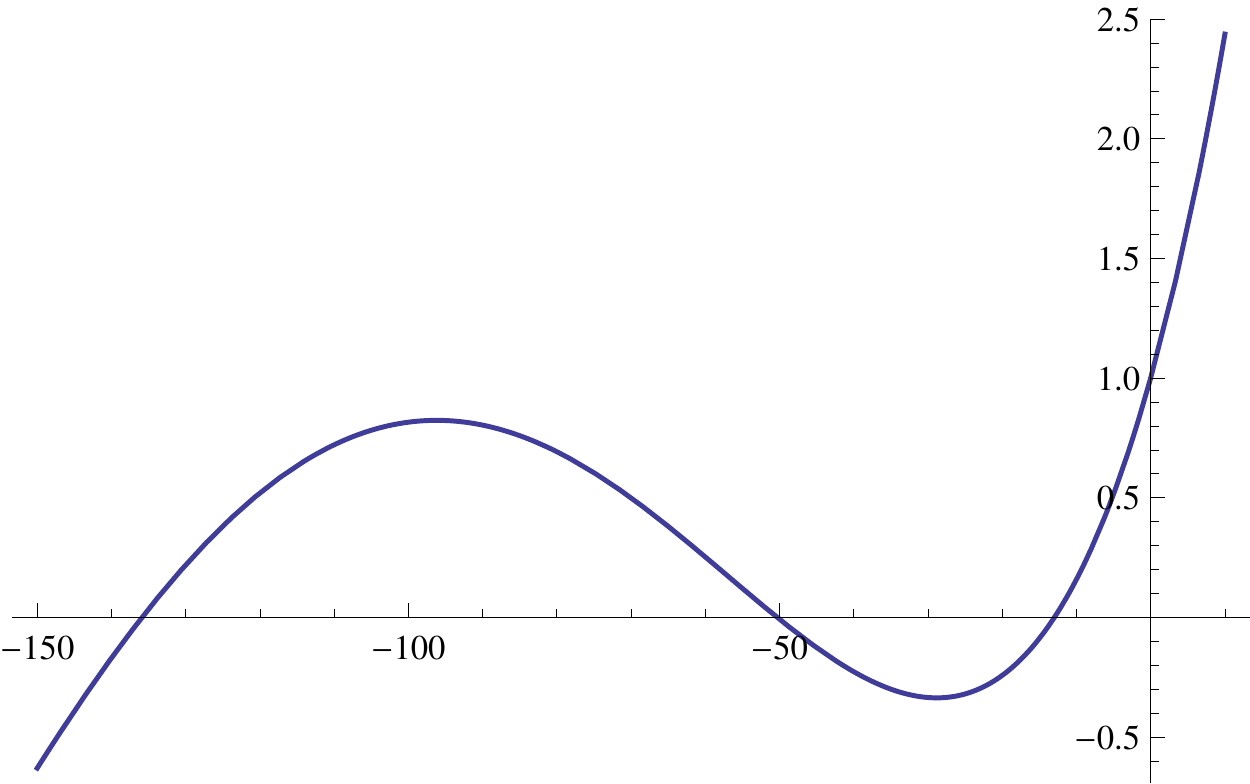}
\caption{The Fredholm determinant $\Xi_{\IP^2} (\kappa, 2 \pi)$ as a function of $\kappa$, showing the first three zeroes on the negative real axis, corresponding 
to the first three energy levels $-\re^{E_n}$, $n=0,1,2$.}
\label{sd-p2-fig}
\end{figure}

\begin{example} Local $\IP^2$ has one single K\"ahler parameter, and no mass 
parameters. The function $A(\hbar)$ appearing in (\ref{jm2}) has been conjectured in closed form, and it is given by 
\be
\label{ah-p2}
A(\hbar) ={3 A_{\rm c}(\hbar/\pi)- A_{\rm c}(3\hbar/\pi) \over 4}, 
\ee
where
 \be
\label{ak}
A_{\rm c}(k)= \frac{2\zeta(3)}{\pi^2 k}\left(1-\frac{k^3}{16}\right)
+\frac{k^2}{\pi^2} \int_0^\infty \frac{x}{\re^{k x}-1}\log(1-\re^{-2x})\rd x.
\ee
This function was first introduced in \cite{MP}, and determined in integral form in \cite{HHHNSY,HO}. It can be obtained by an appropriate 
all-genus resummation of the constants $C_g$ appearing in (\ref{genus-g}). 
In the ``maximally supersymmetric case" $\hbar=2 \pi$, one can write the Fredholm determinant in closed form. 
The standard topological string genus zero free energy is given in (\ref{f0-def}), (\ref{f0-ex}). The genus one free energies are given by \cite{KZ,HK}:
\be
\ba
F_1(t)&={1\over 2} \log\left(-{\rd z \over \rd t}  \right)-{1\over 12} \log\left( z^7 \left(1+27 z \right) \right), \\
F_1^{\rm NS}(t)&= -{1\over 24} \log \left({1+27 z \over z}\right). 
\ea
\ee
We identify 
\be
z=\re^{-3 \mu}={1\over \kappa^3}. 
\ee
Then, it follows from the conjecture above that
\be
\ba
\Xi_{\IP^2}(-\kappa, 2 \pi)&= {\rm det}\left(1-\kappa \rho_{\IP^2} \right)\\
&=\exp \left\{ A(2\pi)+
  {1\over 4 \pi^2} \left( F_0(t) - t \partial_t F_0(t) + {t^2 \over 2} \partial_t^2 F_0(t)\right)+ F_1(t)+F_1^{\rm NS}(t)\right\} \\
  & \times \re^{ \pi \ri/8} \vartheta_2 \left(\xi -{1\over 4}, \tau \right), 
  \label{sd-p2}
  \ea
\ee
where 
 \be
 \label{xi-f}
 \xi= {3\over 4 \pi^2} \left( t \partial_t^2 F_0(t) -\partial_t F_0(t)\right),  \qquad 
 \tau={2 \ri \over \pi} \partial_t^2 F_0(t), 
 \ee
and $ \vartheta_2 \left(z, \tau \right)$ is the Jacobi theta function. The $\tau$ appearing here is the standard modulus 
of the genus one mirror curve of local $\IP^2$. In particular, one has that ${\rm Im}(\tau)>0$. 
It can be easily checked that the function $\xi$ agrees with (\ref{xie}), with the identification $\mu=E$. 
The formulae that we have written down here are slightly different from the ones listed earlier in this section, 
since we are changing the sign of $\kappa$ in the Fredholm determinant, but they can be easily derived from them. We can now perform a 
power series expansion around $\kappa=0$, by using analytic continuation of the topological string 
free energies to the so-called {\it orbifold point}, which corresponds to the limit $z\rightarrow \infty$. This is a standard exercise 
in topological string theory (see for example \cite{small-distances,abk}), and one finds \cite{GHM}
\be
\label{xi-exp}
\Xi_{\IP^2} (\kappa, 2 \pi)=1 + {\kappa \over 9} + \left( {1\over 12 {\sqrt{3}} \pi}-{1\over 81} \right) \kappa^2 + \CO(\kappa^3), 
\ee
provided some non-trivial identities are used for theta functions. This predicts the values of the very first fermionic spectral traces $Z_{\IP^2}(N, 2\pi)$, for $N=1,2$. Interestingly, 
these traces can be calculated directly in spectral theory, since $\rho_{\IP^2}$ is the operator $\rho_{1,1}$ defined in (\ref{rhomn}). 
The integral kernel of this operator is given in (\ref{ex-k}) (for $m=n=1$), and the values 
appearing in the expansion (\ref{xi-exp}) have been verified in this way in \cite{KM,OZ}. 

Finally, it can be easily checked that the function (\ref{sd-p2}) vanishes precisely at the values of $\mu$ given by the quantization condition (\ref{p2qc}) (since we are changing 
the sign of $\kappa$ in this formula, the zeros occur at $\kappa=\re^{E_n}$, $n=0,1, \cdots$). Note that this corresponds to the vanishing locus of the Jacobi 
theta function appearing in (\ref{sd-p2}). In \figref{sd-p2-fig} 
we show a plot of the Fredholm determinant $\Xi_{\IP^2} (\kappa, 2 \pi)$ as a function of $\kappa$ on the real axis. 
For large, positive values of $\kappa$, its behavior is dictated by the large $\mu$ behavior of $\mJ_{\IP^2}(\mu, 2 \pi)$, i.e. 
\be
\Xi_{\IP^2} (\kappa, 2 \pi)\approx \exp\left\{  {3 \over 8 \pi^2} \left( \log(\kappa) \right)^3 \right\}, \qquad \kappa \gg 1. 
\ee
On the other hand, for large, negative $\kappa$, one has the oscillatory behavior, 
\be
\Xi_{\IP^2} (-\kappa, 2 \pi)\approx \exp\left\{  {3 \over 8 \pi^2} \left( \log(\kappa)\right)^3  \right\}  \cos \left( \pi \left( \xi-{1\over 4} \right) \right), \qquad \kappa \gg 1, 
\ee
where $\xi$ is again given by (\ref{xi-f}). \qed \end{example}

What happens in the general case, when $\hbar \not= 2 \pi$? One can of course still compute the Fredholm determinant from our conjecture (\ref{spec-det}), 
but it now contains contributions from all the higher genus Gopakumar--Vafa invariants, and from the all-orders NS free energy. In order to study its properties, one possibility is 
to expand it in power series around the ``semiclassical" case $\hbar=2 \pi$. This was done in \cite{Gr}, and the resulting expansion puts on a rigorous footing the $1/N$ 
expansion of the ``non-perturbative partition function" studied in \cite{EM}. One can also use (\ref{spec-det}) to determine the spectrum by looking at 
the vanishing locus of the Fredholm determinant for general $\hbar$. By using the large radius expansion of the grand potential, one can write an 
exact quantization condition defined by an explicit power series. As shown in \cite{GHM} in many examples, the predictions for the spectrum obtained 
from this exact quantization condition are in perfect agreement with direct numerical calculations (as done for example in \cite{HW}). 
It was found in \cite{wzh} that, when the mirror curve has genus one, the quantization condition of \cite{GHM} 
can be written down in closed form in terms solely of the NS free energy. The results of \cite{wzh,GHM}, when put together, imply 
that the generating functional (\ref{NS-j}), capturing the NS limit of the topological string, is closely related to the generating functional (\ref{GVgf}), capturing the 
{\it standard} topological string. This connection has been further explored in \cite{SWH} and finally established in \cite{GG} in many cases.

For a generic value of $\hbar$, it is difficult to extract analytic results for the spectral traces from (\ref{sd-p2-fig}), since the analytic continuation 
of the generating functionals (\ref{GVgf}), (\ref{NS-j}) to the orbifold point for arbitrary $\hbar$ is not known. 
However, one can still obtain very precise {\it numerical} 
results. The basis for this is the formula (\ref{int-Z}). By expanding $\mJ_S(\mu, \boldsymbol{\xi}, \hbar)$ at large $\mu$, 
the fermionic spectral trace $Z_S(N, \hbar)$ can be evaluated as an infinite sum of Airy functions and their derivatives. Let us spell this out in some 
detail for local $\IP^2$. One can write, 
\begin{equation}
\re^{\mJ_{\IP^2}(\mu, \hbar)} = \re^{\mJ^{({\rm p})}_{\IP^2} (\mu,  \hbar)} \sum_{l, n} a_{l,n} \mu^n \re^{- l \mu}, 
\end{equation}
where
\be
\mJ^{({\rm p})}_{\IP^2} (\mu,  \hbar)={C(\hbar) \over 3} \mu^3 + B(\hbar) \mu+ A(\hbar). 
\ee
The values of $C(\hbar)$, $B(\hbar)$ can be read off from the general expression (\ref{jtotal}) and are given by \cite{GHM}
\be
\label{cb-p2}
C(\hbar)= {9 \over 4 \pi \hbar}, \qquad B(\hbar) ={\pi \over 2 \hbar}-{\hbar \over 16 \pi}. 
\ee
Then, assuming good convergence properties for the expansion of $\mJ_{\IP^2}(\mu, \hbar)$, we can exchange it with integration in (\ref{int-Z}). 
We obtain in this way, 
\begin{equation}
\label{z-exp}
Z_{\IP^2}(N, \hbar) =\frac{\re^{A(\hbar)}}{\left(C(\hbar) \right)^{1/3}}
\sum_{l,n} a_{l,n} \left(-\frac{\partial}{\partial N}\right)^n \mathrm{Ai}
\left(\frac{N+l -B(\hbar)}{\left(C(\hbar)\right)^{1/3}}\right), 
\end{equation}
where ${\rm Ai}(z)$ is the Airy function. Note that $n$ takes non-negative integer values, but $l$ is of the form $3 p + 6 \pi q/\hbar$, where 
$p,q$ are non-negative integers. The leading behavior of (\ref{z-exp}) in the limit $N \rightarrow \infty$, $\hbar$ fixed, is given by the first term in the r.h.s., which is an Airy function 
\be
\label{ai-z}
Z_{\IP^2}(N, \hbar) \approx {\rm Ai} \left(\frac{N -B(\hbar)}{\left(C(\hbar)\right)^{1/3}}\right), \qquad  N \gg 1.
\ee
The additional terms in the r.h.s. of (\ref{z-exp}) give an infinite series of exponentially suppressed corrections to (\ref{ai-z}). Remarkably, this series seems to be 
convergent, and it should be regarded as 
the analogue in this theory of the Rademacher expansion in number theory. 
It also produces highly accurate numerical answers for the spectral traces $Z_{\IP^2}(N, \hbar)$. 
Using this procedure, one obtains for example, 
\be
Z_{\IP^2}\left(1, \hbar={2 \pi \over 3} \right)= 0.4604521481728325977904889856168747087632124207...
\ee
This can be compared to the analytic result obtained by integrating (\ref{ex-k}), 
\be
\tr \rho_{1,1}\left(\hbar ={2 \pi \over 3} \right)= {1\over 3} \exp \left( {V \over 2 \pi} \right), \qquad V= 2 \, {\rm Im}\left( {\rm Li}_2 \left(\re^{\pi \ri / 3} \right)  \right). 
\ee
The agreement is quite remarkable, and the precision increases with the number of terms retained in the expansion of the grand potential. 
Many tests of the main conjecture (\ref{spec-det}), for various local del Pezzo CYs, can be found in \cite{GKMR}. In all cases, one finds that (\ref{spec-det}) captures 
in exquisite detail the spectrum and the spectral traces of the corresponding operator. 

It follows from (\ref{ai-z}) that, at large $N$ and $\hbar$ fixed, the spectral traces display the $N^{3/2}$ scaling
\be
\label{n32}
\log Z_{\IP^2}(N, \hbar)\approx  -{4 {\sqrt{\pi}} \over 9} N^{3/2}\hbar^{1/2}, \qquad  N \gg 1. 
\ee
This can be also derived as a consequence of the leading WKB scaling (\ref{bs}), see for example \cite{MP} for a detailed 
explanation of this type of relation. The above behaviors, (\ref{ai-z}) and (\ref{n32}), are not restricted to local $\IP^2$, but hold 
for all local del Pezzo CYs: the leading term in the large $N$ expansion of the fermionic spectral trace is always an Airy function of the form (\ref{ai-z}), 
which leads to the $N^{3/2}$ scaling. These behaviors are typical of a theory of $N$ M2 branes \cite{kt}, and they are very similar to what happens in ABJM theory \cite{dmp,fhm,MP}. A universal Airy behavior for topological strings has been recently discovered in \cite{ayz, couso}, and it 
would be interesting to understand its precise relation to what is found in this formalism. 

\section{From spectral theory to topological strings} 
\label{ts-st}
\subsection{Random matrices from operators and topological strings} An important consequence of the conjecture of \cite{GHM} is the following. As we have seen, the Fredholm determinant 
involves both the WKB grand potential, which is captured by the NS free energy, and the worldsheet instanton grand potential, which is essentially given by the topological string free energy. 
It turns out that, due to their functional form, there are appropriate scaling limits in which only one of these ingredients contributes. We will focus on the 't Hooft-like limit
\be
\label{thooft-mu}
 \hbar \rightarrow \infty, \qquad \mu \rightarrow \infty, \qquad {\mu \over \hbar} =\zeta \, \, \, \text{fixed}. 
 \ee
 For simplicity, we will also assume that the mass parameters $\xi_j$ scale in such a way that
 \be
 m_j=\xi_j^{2\pi/\hbar}, \qquad j=1, \cdots, s-1, 
 \ee
are fixed (although other limits are possible, see \cite{mz} for a detailed discussion). 
It is easy to see that, in this limit, the quantum mirror map becomes trivial, and the approximation (\ref{tmu}) is exact. The grand potential has
then the asymptotic expansion, 
\be
\label{th-J}
\mJ^{\text{'t Hooft}}_S\left(\zeta, {\boldsymbol{m}}, \hbar \right)= \sum_{g=0}^\infty \mJ^S_g \left(\zeta, {\boldsymbol{m}}\right) \hbar^{2-2g}, 
\ee
where
\be 
\label{gen-J-as}
\ba
 \mJ^S_0 \left( \zeta, {\boldsymbol{m}}\right)&={1\over 16 \pi^4} \left(
 \widehat F_0 \left( {\boldsymbol t}\right) + 4 \pi^2 \sum_{i=1}^{s} b_i^{\rm NS} t_i
  + 14 \pi^4 A_0 \left({\boldsymbol m}\right) \right), \\
 \mJ^S_1 \left(  \zeta, {\boldsymbol{m}}\right)&=  A_1\left({\boldsymbol m}\right)+  \widehat F_1 \left( {\boldsymbol t}\right), \\
  \mJ^S_g \left( \zeta, {\boldsymbol{m}} \right)&=  A_g \left({\boldsymbol m}\right)+ (4 \pi^2)^{2g-2} \left( \widehat F_g \left( {\boldsymbol t}\right) -C_g \right), \qquad g\ge 2. 
  \ea
  \ee
In these equations, 
\be
\label{T-zeta}
 t_i = 2 \pi c_i \zeta-\sum_{j=1}^{s-1} \alpha_{ij} \log m_j, 
\ee
and  $ \widehat F_g \left( {\boldsymbol t}\right)$ are the standard topological string free energies as a 
function of the standard K\"ahler parameters ${\boldsymbol t}$, after turning on 
the B-field as in (\ref{K-sign}). We have also assumed that the function $A\left({\boldsymbol{\xi}}, \hbar\right)$ has the expansion 
\be
A\left({\boldsymbol{\xi}}, \hbar\right)= \sum_{g=0}^\infty A_g({\boldsymbol m}) \hbar^{2-2g}, 
\ee
and this assumption can be tested in examples. 

Therefore, in the 't Hooft limit (\ref{thooft-mu}), the grand potential has an asymptotic expansion which is essentially equivalent to the 
genus expansion of the standard topological string. According to our conjecture, this expansion gives a concrete {\it prediction} for the 't Hooft expansion 
of the fermionic spectral traces, thanks to the equality (\ref{int-Z}). Indeed, let us consider 
the corresponding 't Hooft limit of the fermionic spectral traces, 
\be
\label{thooft-N}
 \hbar \rightarrow \infty, \qquad N \rightarrow \infty, \qquad {N \over \hbar} =\lambda \, \, \, \text{fixed}. 
 \ee
Then, it follows from (\ref{int-Z}) that one has the following asymptotic expansion, 
\be
\label{zn-asex}
\log\, Z_S(N, \hbar) = \sum_{g\ge 0} \CF_g(\lambda) \hbar^{2-2g}, 
\ee
where $\CF_g(\lambda)$ can be obtained by evaluating the integral in the right hand side of (\ref{int-Z}) in the saddle-point approximation and using just the 
't Hooft expansion of $\mJ_S(\mu, \boldsymbol{\xi}, \hbar)$. At leading order in $\hbar$, this is just a Legendre transform, and one finds 
\be
\label{legendre}
\CF_0(\lambda)=\mJ^S_0\left(\zeta, {\boldsymbol{m}} \right)- \lambda \zeta, 
\ee
evaluated at the saddle-point given by 
\be
\lambda={\partial  \mJ^S_0 \over \partial \zeta}. 
\ee
The higher order corrections can be computed systematically. In fact, the formalism of \cite{abk} gives a nice geometric description of the integral transform (\ref{int-Z}) in the 
saddle-point approximation: the functions $\CF_g(\lambda)$ are simply the free energies of the topological string on $X$, but on a different frame. This frame corresponds to 
the so-called {\it conifold} point of the geometry. In particular, $\lambda$ turns out to be a vanishing period at the conifold point.

Therefore, the conjecture (\ref{spec-det}), in its form (\ref{int-Z}), provides a precise prediction for the 't Hooft limit of the fermionic spectral traces: they are encoded in the standard 
topological string free energy, but evaluated at the conifold frame. It turns out that this prediction can be tested in detail. The reason is that, at least in some cases, the fermionic 
spectral traces can be expressed as random matrix integrals, and their 't Hooft limit can be studied by using various techniques developed for matrix models at large $N$. 
In the case of the three-term operators (\ref{tto}), the explicit expression for the kernel appearing in (\ref{ex-k}) can be combined with Fredholm's theorem (\ref{fred}) to obtain 
the following expression for their fermionic spectral traces \cite{mz}, 
\be
\label{zmn-bis}
Z_{m,n}(N,\hbar)=\frac{1}{N!}  \int_{\IR^N}  { \rd^N u \over (2 \pi)^N}  \prod_{i=1}^N  \left| \mypsi{a}{c}\left(  {\mb u_i \over 2 \pi} \right) \right|^2 
 \frac{\prod_{i<j} 4 \sinh \left( {u_i-u_j \over 2} \right)^2}{\prod_{i,j} 2 \cosh \left( {u_i -u_j \over 2} + \ri \pi C_{m,n} \right)}. 
\ee
This can be regarded as an $O(2)$ matrix model \cite{kostov,ek}, with a potential depending on $\hbar$ and given by 
\be
V(u, \hbar)=-\hbar \log \left| \mypsi{a}{c}\left(  {\mb u \over 2 \pi} \right) \right|^2. 
\ee
In \cite{mz} it is shown in detail that this potential admits an asymptotic expansion at large $\hbar$, 
which can be used to obtain the 't Hooft expansion of $Z_{m,n}(N,\hbar)$. In particular, 
when $m=n=1$, one finds a description of the all-genus topological string free energy of local $\IP^2$ 
as the asymptotic 't Hooft expansion of this matrix integral. Detailed calculations 
show that the expansion obtained in this way agrees exactly with the predictions of (\ref{int-Z}). In fact, the 
conjecture ``explains" many aspects of topological string theory at the conifold point, 
for many toric CY geometries. For example, it has been known for some time that the leading terms in the expansion of the 
conifold free energies are universal \cite{gv-coni}, and that they agree with the all-genus free energy of the 
Gaussian matrix model. Since matrix models like (\ref{zmn-bis}) are deformations of the Gaussian one, this observation is 
a consequence of our conjecture (or, equivalently, this observation 
must hold in order for the conjecture to be true). 
\begin{figure}[h]
\center
\includegraphics[scale=0.375]{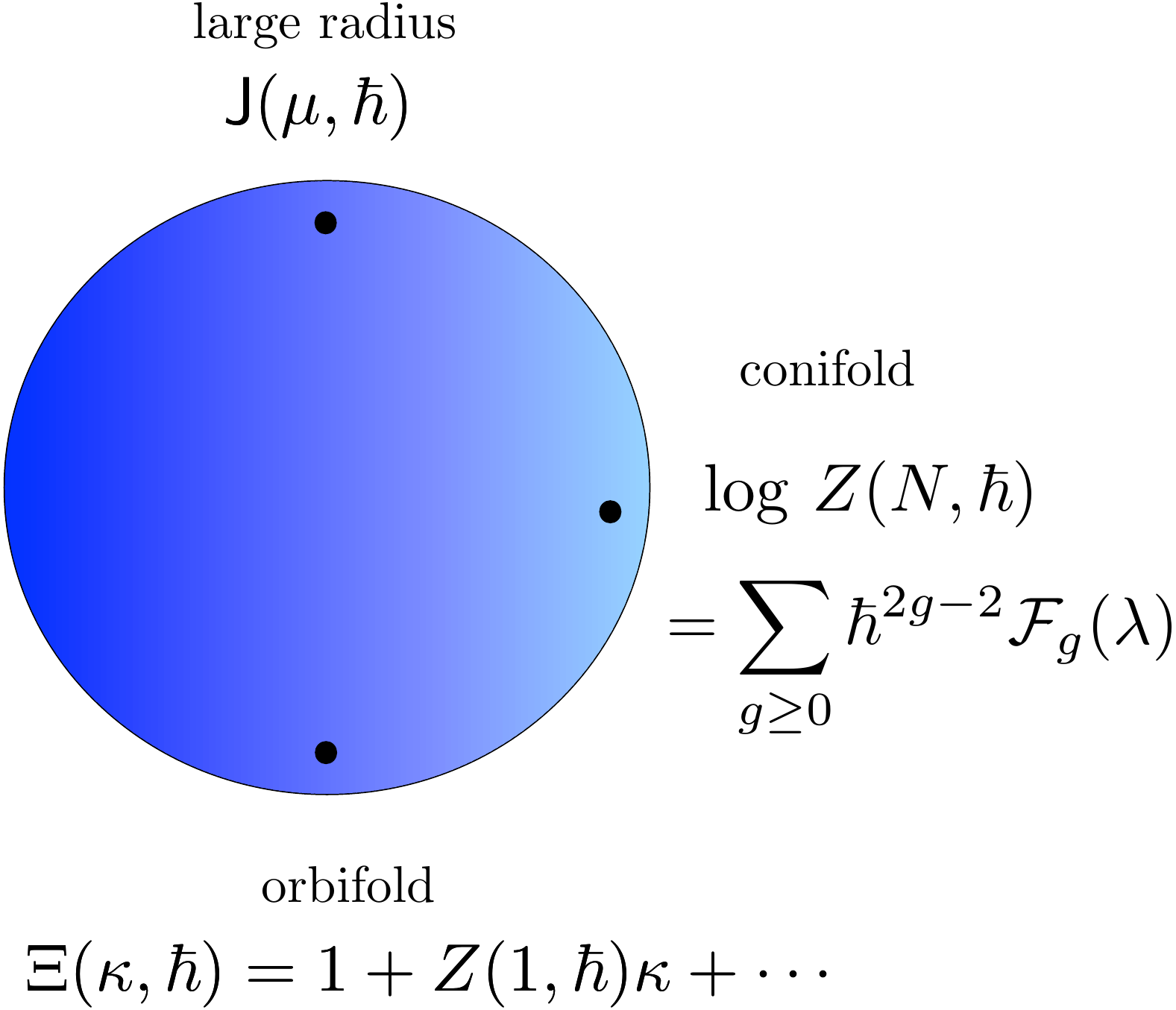}
\caption{Different points in the moduli space lead to different expansions for the spectral quantities.}
\label{moduli-dif}
\end{figure}

Note that, as shown in \figref{moduli-dif}, different regions of the moduli space of a toric CY are 
associated to different expansions of the spectral quantities: the large radius region defines the grand potential at large $\mu$. The orbifold 
point is the appropriate one to calculate the Taylor expansion of the Fredholm determinant at the origin, as in (\ref{xi-exp}), while the conifold point encodes 
the 't Hooft limit of the fermionic spectral traces. 

\subsection{Non-perturbative topological strings}  As we mentioned above, 
the total free energy of the topological string (\ref{tfe}) is hopelessly divergent. This is also the case for the total free energy at the conifold frame. 
Let us consider the formal power series in $\hbar$ appearing in the r.h.s. of 
(\ref{zn-asex}). As it happens at large radius, the functions $\CF_g(\lambda)$ have a common, finite radius of convergence. For each $\lambda$ inside 
this radius of convergence, the resulting series 
in $\hbar$ diverges factorially, like 
\be
\CF_g(\lambda) \sim (2g)! (A(\lambda))^{-2g}, \qquad g\gg 1. 
\ee
Therefore, the expansion (\ref{zn-asex}) is asymptotic and does not define a function of $\lambda$, $\hbar$. However, 
we have now found a quantity, the fermionic spectral trace, with the following properties:
\begin{enumerate}

\item It is rigorously well-defined. This is a consequence of the trace class property of $\rho_S$. 

\item Its asymptotic expansion, in the 't Hooft limit, reproduces exactly the genus expansion of the topological string in the conifold frame.
\end{enumerate}

\begin{figure}[h]
\center
\includegraphics[scale=0.35]{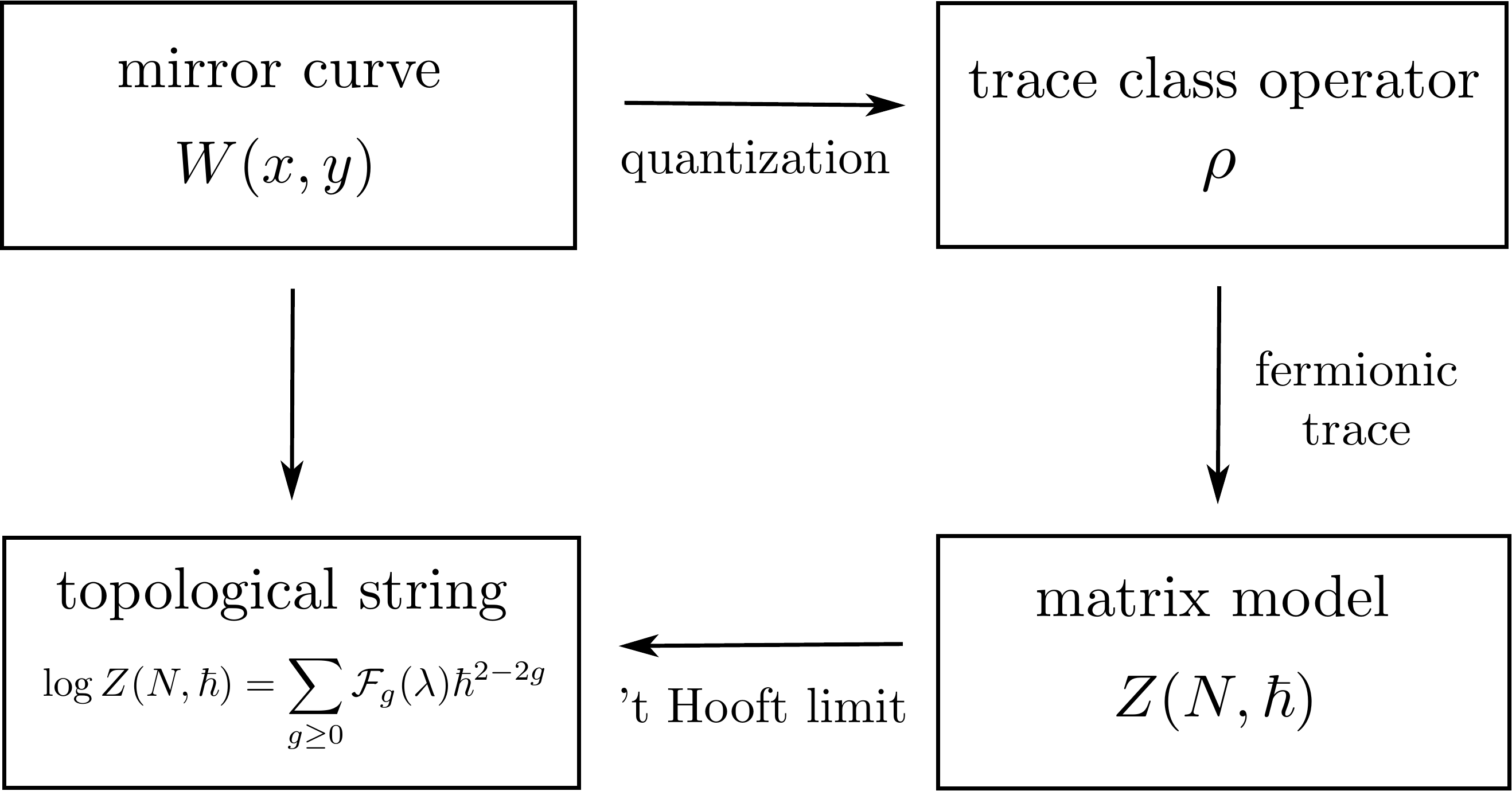}
\caption{Given a toric CY threefold, the quantization of its mirror curve leads to a trace class operator $\rho$. 
The standard topological string free energy is obtained in the 't Hooft limit of its fermionic traces $Z(N, \hbar)$.}
\label{diagram}
\end{figure}

It follows that {\it the fermionic spectral traces provide a non-perturbative completion of the topological string}, in the conifold frame.
The results of \cite{GHM,mz,KMZ} lead then to an elegant approach to non-perturbative topological strings: 
given a toric CY manifold $X$, one considers the quantization of its mirror curve, which leads to a trace class 
operator. The standard, all-genus topological string free energy emerges as an asymptotic expansion of the 
fermionic spectral traces of the operator, in a 't Hooft-like limit. This procedure is sketched in \figref{diagram}. Note that, from the point of view of the spectral problem, 
the conventional topological string emerges at {\it strong} coupling $\hbar \gg 1$. 

Non-perturbative completions are not unique. One way to obtain well-defined answers from asymptotic series is to use Borel 
resummation. In favorable cases, when the series is Borel summable, this procedure is unambiguous and leads to a function whose asymptotic expansion reproduces the original 
asymptotic series (see for example \cite{mm-nprev} for a review). In the case of topological string theory in the conifold frame, it turns out that, in many examples (like local $\IP^2$), the genus expansion is 
Borel summable for real $\hbar$ and real $\lambda$. 
The result of the Borel resummation however turns out to be {\it different} from the non-perturbative completion 
defined by the fermionic spectral traces. For example, in the case of local $\IP^2$, one finds that \cite{CMS}
\be
\log Z^{\rm Borel}_{\IP^2}(1, 2 \pi)=-2.197217... 
\ee
to be compared with the value obtained from spectral theory (see (\ref{xi-exp})),
\be
\log Z_{\IP^2}(1, 2 \pi)=-\log(9)=-2.197224...
\ee
The difference between the Borel resummation and the answer from spectral theory is due to exponentially small, non-perturbative effects associated to complex 
instantons. In the context of the theory of resurgence, this indicates that the perturbative genus expansion has to be generalized to a {\it trans-series} including 
the non-perturbative effects in the form of additional asymptotic series. Such trans-series can be produced by using the powerful techniques of \cite{cesv}, 
which are based on a generalization of the holomorphic 
anomaly equation of \cite{BCOV} to the non-perturbative realm. One can now 
wonder whether the non-perturbative information contained in the fermionic spectral traces can be recovered from the Borel--\'Ecalle resummation 
of the full trans-series. In \cite{CMS}, strong numerical evidence has been given that this is the case \footnote{Non-perturbative effects are 
explicitly included in the NS contribution to the grand potential, but this contribution does not have the 
standard form of a trans-series in the string coupling constant.}, relating in this way the ``exact" non-perturbative approach of \cite{GHM} and 
the resurgent analysis of non-perturbative effects in \cite{cesv}. 

The fermionic spectral trace $Z(N, \hbar)$ provides a non-perturbative completion of the 
topological string in the conifold frame, and it actually promotes it to an {\it entire} function of 
$N$, as first noted in \cite{CGM}. In the large radius frame, we can also promote the topological string partition function to a entire function on moduli space, 
which is nothing but the spectral determinant itself. More precisely, in the 't Hooft regime (\ref{thooft-mu}), we have the asymptotic behavior 
\be
\label{logxi}
\log \Xi_S(\kappa, \boldsymbol{\xi}, \hbar)\sim \mJ^{\text{'t Hooft}}_S(\bf{t}). 
\ee
In this equation, the $t_i$ are related to $\zeta$, $\boldsymbol{m}$ as spelled out in (\ref{T-zeta}). The r.h.s. is given in (\ref{th-J}) and (\ref{gen-J-as}), and it involves the 
genus expansion of the topological string in the large radius regime. Therefore, we can promote the total topological string partition function in the large radius frame, 
to an entire function of moduli space. 
It was pointed out by Witten in \cite{W} that this partition function should be thought as a wavefunction on the moduli space 
of the CY. In this view, the genus expansion should be regarded as a sort of WKB approximation to this wavefunction. However, the WKB solution
has branch cuts and singularities which are an artifact of the approximation, and are not present in the exact answer, which is sometimes 
an entire function. In \cite{mmss}, such a phenomenon has been argued to occur as well in string theory, where perturbative wavefunctions 
should be ``smoothed out" by non-perturbative effects. It is tempting to interpret (\ref{logxi}) in this light, and to regard the 
spectral determinant as the {\it exact}, {\it entire} 
wavefunction on moduli space postulated in \cite{W}. As in the WKB approximation, 
the branch cuts and conifold singularities appearing in the perturbative free energies of the 
topological strings should be regarded as artifacts of the genus expansion, and they are smoothed out in the final answer. 

\subsection{Number-theoretic aspects} It has been observed in \cite{rv,dk} (see also \cite{yang}) that the values of the K\"ahler parameters of many toric CYs at the 
conifold point are given by special values of the dilogarithm function. It turns out that these number-theoretic properties of the CY periods at the conifold point 
are consequences of the main conjecture (\ref{spec-det}). 
According to this conjecture (\ref{spec-det}), the topological string amplitudes are encoded in the spectral properties of the 
operator obtained by quantizing the curve. On the other hand, 
as shown in \cite{KM}, this operator is closely related to the quantum dilogarithm. When these two statements are put together, the results of \cite{rv,dk} follow.

To see this in more detail, note that (\ref{legendre}) implies that
\be
{\partial \CF_0 \over \partial \lambda}= -\zeta. 
\ee
It follows from (\ref{zn-asex}) that the l.h.s. of this equation can be calculated by 
studying the large $N$ limit of the fermionic spectral traces. For the operators $\mO_{m,n}$, it can be computed from the matrix integral 
(\ref{zmn-bis}), and the result is \cite{mz}
\be
\label{bw-eval}
\left({\partial \CF_0 \over \partial \lambda}\right)_{\lambda=0}= \frac{m+n+1}{2\pi^2} D(-q^{m+1}\chi_m).
\ee
In this equation, 
\begin{align}
q=\exp\left(\frac{\ri \pi}{m+n+1}\right), \qquad \qquad \chi_k=\frac{q^k-q^{-k}}{q-q^{-1}}, 
\end{align}
and $D(z)$ is the Bloch--Wigner function defined by, 
\be
D(z)={\rm Im \, Li_2}(z)+{\rm arg}(1-z)\log|z|, 
\ee
where arg denotes the branch of the argument between $-\pi$ and $\pi$. Now, the point $\lambda=0$ is the conifold point, 
and since $\zeta$ is related to the K\"ahler parameters $t_i$ by (\ref{T-zeta}), one can evaluate these parameters at the conifold 
point in terms of (\ref{bw-eval}). 

The simplest example of this situation is again local $\IP^2$. In this case there is a single K\"ahler parameter $t$, related to $\zeta$ by 
\be
t= 6 \pi \zeta. 
\ee
The conifold point occurs at $z=-1/27$, which is at the boundary of convergence of the series in (\ref{pers-2}). 
The value of $t$ at this point, which we will denote by $t_c$, can be in principle 
obtained by 
evaluating the mirror map (\ref{lp2-mm}), 
\be
\label{tc-con}
t_c= -\log(1/27) - \widetilde \varpi_1\left( -1/27 \right).  
\ee
Here, we are dropping an imaginary piece $\pm \pi \ri$ coming from the log. According to (\ref{bw-eval}), (\ref{tc-con}) can be evaluated in terms of the Bloch--Wigner function as 
\be
\label{tc-bw}
t_c={9 \over \pi} D\left( \re^{\pi \ri /3} \right). 
\ee
This is clearly a non-trivial number-theoretic property of the period. In the framework of \cite{GHM,mz}, it follows ultimately from the fact that 
the operator $\rho_{\IP^2}$ involves the quantum dilogarithm \cite{KM}. The result (\ref{tc-bw}) 
turns out to be true, and it has been proved in \cite{rv,dk}. Similar identities can be obtained for other CYs, and some of them seem to be new (see \cite{cgm2,CGuM} for examples involving 
 mirror curves of genus two.) 

The fact that the spectral theory realization of the topological string ``explains" some number-theoretic properties of the periods seems to indicate that this 
non-perturbative completion is capturing essential aspects of the theory.

 \section{Outlook}
\label{out}
In this paper we have reviewed the correspondence between spectral theory and topological 
strings presented in \cite{GHM}. This correspondence leads to a new family of exactly solvable 
problems in operator theory, providing explicit expressions for the corresponding Fredholm determinants. 
At the same time, it leads to new insights 
on topological strings on toric CY threefolds, and to a non-perturbative definition of these string 
theories in the spirit of large $N$ string/gauge dualities. 

In this review, we have focused on mirror curves of genus one. The case of genus zero is certainly special, since 
the corresponding operators do not have discrete spectra. 
However, the ideas of \cite{hmmo,GHM} can be extended to this case in a natural way, as it has been 
noted recently in \cite{hatsuda,HO2,krefl}. More interesting, and more difficult, 
is the extension to mirror curves of higher genus. This has been done in \cite{cgm2} and further explored in \cite{CGuM}. 
When the mirror curve  has genus $g_\Sigma$, its quantization
leads to $g_\Sigma$ different trace class operators. It is possible to define a generalized 
Fredholm determinant depending on $g_\Sigma$ moduli. There are also natural generalizations of the 
fermionic spectral traces, which now depend on $g_\Sigma$ different non-negative integers $N_i$, 
$i=1, \cdots, g_\Sigma$, and provide a non-perturbative definition of the topological string in a 't Hooft-like limit. However, in contrast to the situation for 
quantum integrable systems, there is a single quantization condition, which determines a codimension one locus in moduli space through the vanishing 
of the Fredholm determinant

The results presented here have important consequences for the study of a class of integrable systems, called {\it cluster integrable systems}, which can be 
associated to local CY manifolds. These systems were constructed by Goncharov and Kenyon in \cite{GoKe} and include in particular 
the relativistic Toda lattice \cite{efs,marshakov,fm}. The number of Hamiltonians in these systems is equal to the genus of the mirror curve. In the case of curves of genus one, the operator $\mathsf{O}_S$ is 
nothing but the (single) quantum Hamiltonian of the Goncharov--Kenyon system. Therefore, the results of \cite{GHM} provide a conjectural solution to the 
quantum integrable system in the genus one case. By using the formulation 
of the quantization condition of \cite{GHM} presented in \cite{wzh}, an exact quantization condition for the spectrum of the 
general Goncharov--Kenyon system has been conjectured in \cite{hm,frhama}. 

There are clearly many avenues to explore in the future. Although we have presented some detailed results for the spectrum of the operators, one should 
address the construction of the eigenfunctions, which is closely related to the open string sector. 
A first step in this direction has been made in \cite{MZ2}, where a general proposal has been made in the case of genus one curves, 
and an exact solution has been written down explicitly in the maximally supersymmetric case of local $\IP^1\times \IP^1$. However, much more work 
is needed in order to have a general solution for the eigenfunctions. Eventually, one should 
provide a rigorous proof (or at least a derivation at the physics level of rigor) of the quantization condition. This looks difficult, since, for this type of operators, 
we do not even have a semi-rigorous method to calculate instanton corrections. Developing these tools is an important task for the future.

On the topological string side, one limitation of the framework we have presented is that it does not accommodate the general refined theory, 
but only the two one-parameter specializations corresponding to the standard and the NS limits. However, one lesson 
we have learned from the developments reviewed in this paper is that these two 
one-parameter cases are not independent: in order to solve the spectral problem, we need 
both generating functions, (\ref{GVgf}) and (\ref{NS-j}), and each one of them 
can be regarded as a non-perturbative correction to the other. 
Maybe we should think about the refined topological string with general $(\epsilon_1, \epsilon_2)$ 
parameters as a non-perturbative object, similar to the $(p,q)$ string, instead of 
looking for a perturbative worldsheet description.


%
%


\bibliographystyle{hamsalpha}
\bibliography{biblio}
\end{document}